\documentclass[compsoc,journal]{IEEEtran}

\usepackage{hyperref}
\usepackage{xcolor,colortbl}
\newcommand{\fig}[1]{Figure~\ref{fig:#1}}
\newcommand{\tbl}[1]{Table~\ref{tbl:#1}}
\newcommand{\eq}[1]{Equation~\ref{eq:#1}}

\usepackage{balance}
\usepackage{wrapfig}
\usepackage{multirow}
\usepackage{blindtext, graphicx}
\usepackage{boldline}
\usepackage{subfig}
\usepackage{booktabs}
\usepackage{ragged2e}
\usepackage{dblfloatfix}
\usepackage[para,online,flushleft]{threeparttable}
\newcommand{\IT}[1]{{{\em #1}}}

\usepackage{lipsum}

\usepackage[tikz]{bclogo}
{\noindent\begin{minipage}[c]{\linewidth}%
\begin{bclogo}[couleur=gray!25,%
                arrondi=0.1,%
                logo=\bctrombone,%
                ombre=true]{{\small ~#1}}}%
{\end{bclogo}\end{minipage}\vspace{2mm}}

\newcommand{\bi}{\begin{itemize}}
\newcommand{\ei}{\end{itemize}}

\usepackage{enumitem}
\setlist[itemize]{leftmargin=*}
\setlist[enumerate]{leftmargin=*}
\usepackage{makecell}
\usepackage[linesnumbered,ruled,vlined]{algorithm2e}
\setlist{nolistsep}

\setlist[1]{itemsep=0pt}

\usepackage{amsmath}

\usepackage{color}
\usepackage{listings}
\usepackage{times}
\usepackage{colortbl}
\usepackage{tikz}

\makeatletter
\let\th@plain\relax
\makeatother

\definecolor{Gray}{rgb}{0.88,1,1}
\definecolor{Gray}{gray}{0.85}
\definecolor{lightgray}{gray}{0.8}
\usepackage[framed]{ntheorem}
\usepackage{framed}
\usepackage{tikz}
\usetikzlibrary{shadows}
\theoremclass{Lesson}
\theoremstyle{break}

\tikzstyle{thmbox} = [rectangle, rounded corners, draw=black,
fill=gray!25,  drop shadow={fill=black, opacity=1}]

\newshadedtheorem{lesson}{Answer to RQ}

\definecolor{myGray}{RGB}{220, 220, 220}
\definecolor{myRed}{RGB}{156, 4, 4}
\definecolor{myBlue}{RGB}{108, 156, 236}
\definecolor{myGreen}{RGB}{108, 172, 76}

\usepackage{xcolor}

\usepackage{rotating}
\usepackage{tabularx}
\usepackage{tabu}
\usepackage{adjustbox}

\usepackage[framemethod=tikz]{mdframed}
\usetikzlibrary{shadows}
\usepackage{graphics}
\newmdenv[
tikzsetting= {fill=white},
linewidth=1pt,
roundcorner=2pt, 
shadow=false
]{myshadowbox}

\newenvironment{result}[2]
{\vspace{1mm}
\begin{myshadowbox}\textbf{\textit{\underline{Lesson#1:}}} #2}{ 
\end{myshadowbox}}

\newcommand{\crule}[3][darkgray]{\textcolor{#1}{\rule{#2}{#3}}}
\newcommand{\dbox}[1] { \crule[black!#1]{0.62cm}{0.3cm} 
\hspace{-0.51cm}\scalebox{1}[1.0]{{\textcolor{black}{{\bf $^{#1}$}}}\hspace{0.1mm}}}
\newcommand{\wbox}[1] { \crule[black!#1]{0.62cm}{0.3cm} 
\hspace{-0.51cm}\scalebox{1}[1.0]{{\textcolor{white}{{\bf $^{#1}$}}}\hspace{0.1mm}}}

\graphicspath{{images/}}
\usepackage{rotate}
\usepackage{listings}
\usepackage{colortbl}
\usepackage{arydshln}
\usepackage{bigdelim}
\usepackage{xcolor}
\usepackage{balance}
\usepackage{booktabs}
\usepackage{wrapfig}

\usepackage{tikz}
\usetikzlibrary{angles}
\usepackage{makecell}
\usepackage{multirow}
\usepackage{url}
\usepackage{color}

\bstctlcite{IEEEexample:BSTcontrol}

\begin{document}

\SetKwProg{Fn}{Function}{}{}
%

\author{Tianpei Xia,~
        Rui Shu,~
        Xipeng Shen,
        and Tim Menzies,~\IEEEmembership{Fellow,~IEEE}
\IEEEcompsocitemizethanks{\IEEEcompsocthanksitem T. Xia, R. Shu, X. Shen, and T. Menzies are with the Department
of Computer Science, North Carolina State University, Raleigh, USA.\protect\\
E-mail: \{txia4, rshu, xshen5\}@ncsu.edu, timm@ieee.org}}




\title{Sequential Model Optimization \\ for Software Process Control  }

\IEEEtitleabstractindextext{%
{\justify\begin{abstract}
Many methods have been  proposed to estimate how much effort is required to build and maintain software. Much of that research assumes a ``classic'' waterfall-based approach rather than contemporary projects (where the developing process may be more
iterative than linear in nature). Also, much of that work tries to recommend a single method-- an approach that makes the dubious assumption that  one method can handle the diversity of software project data. 

To address these drawbacks,   we apply a configuration technique called ``ROME''  (Rapid Optimizing Methods for Estimation), which uses sequential model-based optimization (SMO) to find what combination of
effort estimation
techniques works best for a particular data set. We test this method using data from  1161 classic waterfall projects and 120 contemporary projects (from Github). In terms of magnitude of relative error and standardized accuracy, we find that ROME achieves better performance than existing state-of-the-art methods for both classic and contemporary problems.  In addition, we conclude that we should not recommend {\em one} method for estimation. Rather, it is better to  search through a wide range of different  methods  to find what works best for local data.    

To the best of our knowledge, this is the largest effort estimation experiment yet attempted and the only one to test its methods on classic and contemporary projects.

\end{abstract}}
}

\IEEEdisplaynontitleabstractindextext

\maketitle
\section{Introduction}
\label{sect:intro}

Estimating development effort is hard~\cite{kemerer1987empirical}
and incorrect estimates can harm   software projects~\cite{trendowicz2014software,mcconnell2006software,mendes2002further,sommerville2010software}.  When developers are forced to build their software using too few resources, then  the first thing that is usually  jettisoned is the software quality task~\cite{Menzies2008}. Also,
when monitoring for ``project health'',
the managers of large open source distributions will  shun such distressed projects (so that software will  not  get  widely used~\cite{wahyudin2007monitoring}).

Much of the prior work on effort estimation has focuses on classic waterfall projects~\cite{boehm2000cost,benediktsson2003cocomo,shepperd1997estimating,sheta2006estimation}. In classic waterfall estimation, the goal is to get the budget right, before any work starts.  However,    contemporary projects that use continuous
integration have little need
for that kind of estimate.
Instead, what 
contemporary
projects  need to do is  update their
effort estimators (based on recent activity)
in order to generate short-term estimates
of the effort required to complete the immediate tasks to hand. 

In practice, software may be built using
a diverse combination of waterfall and continuous methods (the so-called ``agilefall'' method~\cite{blank:2019}).
As software engineering  development practices   grow more diverse, it becomes less and less likely that any single estimation model will work across all those projects. So instead of recommending a particular estimation model, we say:
\begin{quote}
{\em   To find what works best for  local data,  we need ways to survey  a wide range of different estimation models.}
\end{quote}
For this task, we recommend a new approach called 
``ROME'' (Rapid Optimizing Methods for Estimation), which uses sequential model-based optimization (SMO) to explore possible configurations for an effort estimator.  In that process, the results from exploring a  few configurations are used to guess results across the remaining configurations.   The configuration that yields the best guess (lowest error)
is then actually applied, after which ROME updates its knowledge of what is a good configuration.    

To evaluate ROME, we ask these research questions:

{\bf \em RQ1: Is effort estimation effective for classic waterfall and contemporary projects?} 
When  software can be built
via a  combination of processes,
effort estimation needs to be effective for both
classic waterfall and contemporary continuous integration
projects.  According to Sarro et al., industrial competitive predictions of project effort usually lie within 0.3  and 0.4 of the actual value~\cite{sarro2016multi}. We provide evidence that the performance of our method in classic waterfall and contemporary data sets lies within the currently claimed industrial human-expert-based thresholds, thereby demonstrating that:

 \begin{result}{1}
 Effort estimation is effective on both classic waterfall projects and contemporary projects.
 \end{result}



{\bf \em RQ2: Does ROME have better performance than existing estimation methods?} 
To answer this question, we study    1161 classic  waterfall projects and 120 contemporary projects (from Github).  ROME's performance is compared to some standard effort estimators  as well as two recent prominent  systems: Whigham et al.'s ATML tool from TOSEM'15~\cite{Whigham:2015} as well as Sarro et al's  LP4EE tool from TOMSE'18~\cite{SarroTOSEM2018}.  We find that:

 \begin{result}{2}
 ROME generate better estimates than other methods in most cases.
 \end{result}
 Here, we measured ``best'' using the measures that are standard in the field; i.e. MacDonell''s and Shepperd's standardized accuracy measure~\cite{shepperd2012evaluating} and  the MRE measure used by   other   researchers~\cite{sarro2016multi}.

 \newpage

{\bf RQ3: \em When we have new effort data sets, what configurations  to use for effort estimation tasks?} The tool we call ROME is a combination of sequential  model optimization (SMO) and regression tree learner.   For pragmatic reasons,   practitioners    prefer a simpler rig. Hence we are often asked
if  the   optimizer is required or if, usually, certain configurations generally work well across all data sets. To answer this question,
we  counted  what configurations were selected in the experiments of this paper. In those counts, we can see:
\begin{result}{3}
There is no clear pattern in what configurations  are needed. Hence,  model optimization needs to be repeated for each new data set.
\end{result}


{\bf RQ4: \em When we apply ROME on effort data sets, can it help us to find the most important features of the data?} 
One feature of ROME is that if a feature is not informative, it will be dropped in the generated estimation model. Hence,
when we say ``most important'', we really mean the ``mostly used in our methods''. Looking across our results, we find that certain size features are always used, but always in combination with a wide variety of other features. Hence:
\begin{result}{4}
There are no ``best'' set of effort estimation features since each project uses these features in a different way.
\end{result}

Overall the contributions of this paper are:
\bi
\item A new collection of contemporary project data from 120 Github repositories. Along with the data from 1161 classic waterfall projects, this is the largest effort estimation experiment yet reported to the best of our knowledge.
\item Using those data, our results from
{\bf RQ1} show that   effort estimation works well for classic waterfall projects as well as contemporary projects.
In terms of the practicality of effort estimation research, this is a landmark result since it means that decades of research into effort estimation of classic waterfall projects can now be applied to contemporary software systems.

\item  The results of our experiments clearly deprecate the use of off-the-shelf estimation tools. Based on the lessons of {\bf RQ2}, {\bf RQ3}  and {\bf RQ4}, practitioners should use tools like ROME to  find the features/modeling options that work best for their local data.
\item  We offer a new benchmark in effort estimation and an open source version of ROME. The latter is more of a system contribution than a research contribution. Nevertheless, in terms of support the reproduction and extension of our results, this contribution is useful.
\ei
The rest of this paper is structured as follows.
The next section discusses the history and different methods for effort estimation tasks. This is followed by a description of our experimental data, methods and the results. After that, a discussion section explores open issues with this work.

\section{Background}
\label{sec:back}

Software effort estimation is the procedure to provide
approximate advice on how much human effort is required to plan, design and develop a software project. Usually, this human effort is expressed in terms of hours, days or months of human work. Since software development is a highly dynamic and fluid process, any estimate can only be approximate. Still, doing estimation is necessary since it is important to allocate resources properly in software projects to avoid waste. In some cases, improper allocation of funding can cause a considerable waste of resource and time~\cite{cowing02,germano16,hazrati11,roman16}.

\begin{figure}[!t]
\centerline{\includegraphics[width=0.5\textwidth]{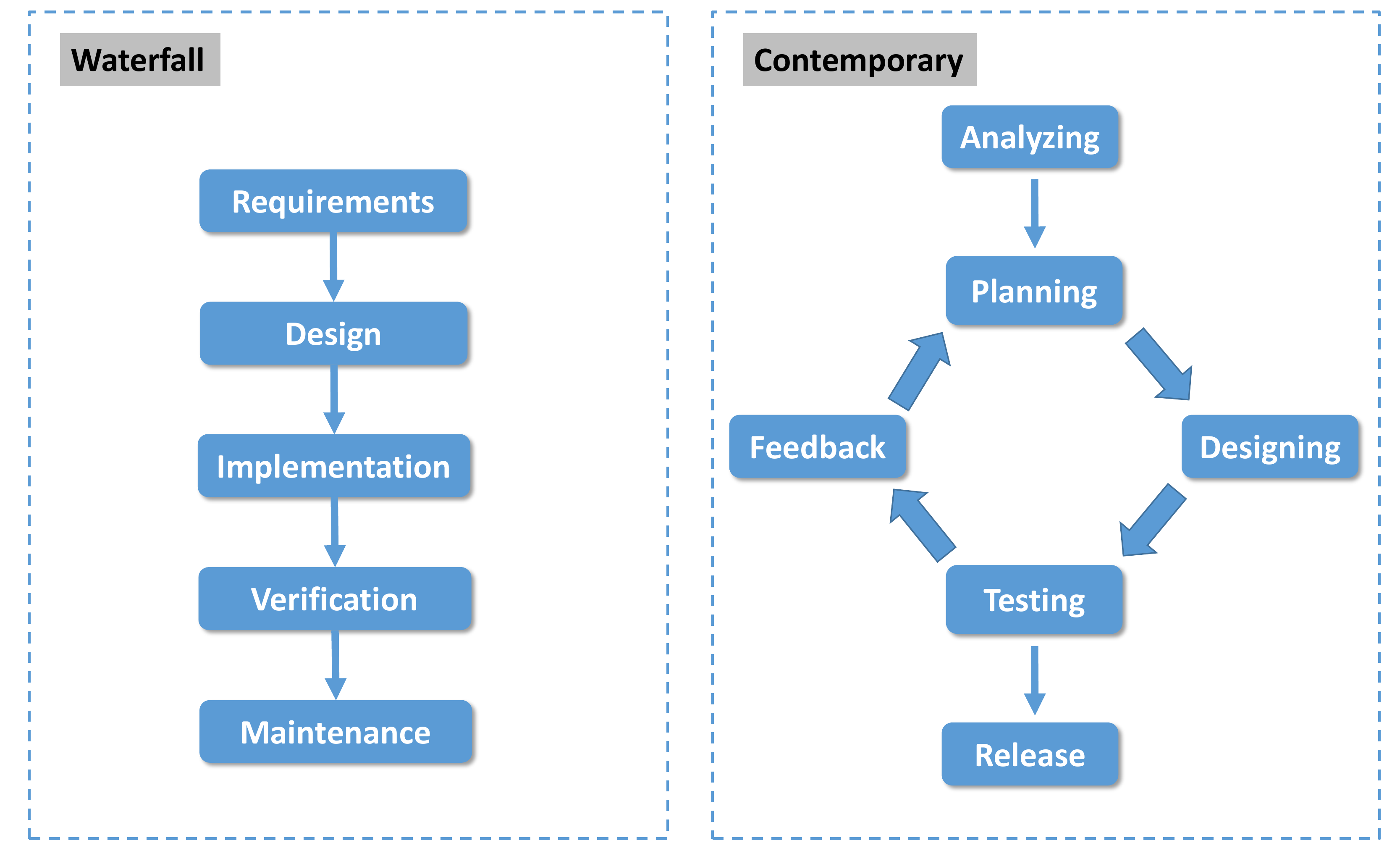}}
\caption{Classic vs. Contemporary styles of  software development.}    
\label{fig:watercon} 
\end{figure}

Much effort estimation work assumes a classic waterfall model~\cite{boehm2000cost,benediktsson2003cocomo,shepperd1997estimating,sheta2006estimation}, first documented   by Royce et al. in1970~\cite{royce1970software}. In this approach, project  teams  move to the next phase of development or testing if the previous step  successfully completes. Estimation  happens before the coding started. Further, once the funds are allocated, there is little opportunity to change that allocation.

Waterfall-style development
is far less common in contemporary
industrial practices. In much of  current   software practice,  development and testing activities are concurrent. Such contemporary methods use a continuous iteration of planning, developing and testing. This allows more communication between customers, developers, managers, and testers. \fig{watercon} contrasts these contemporary and classic kinds of
software development.

Contemporary projects are 
staffed using a 
 fluctuating  population 
 of programmers.
 Hence, whereas classic waterfall projects need estimates  for future work, the managers  of contemporary projects need estimates to know if their current staff allocation is sufficient for the tasks at hand~\cite{robles2014estimating}.
For example, Krishna et al. report that IBM asked for  help to adjust, on a month-to-month basis, the staffing allocations for   their   suite of contemporary open source
tools (which IBM maintains for its client base)~\cite{krishna2018connection}.

Having much of contemporary industrial practice uses continuous deployment, we  hasten to add that waterfall projects still exist. This is particularly true in the case of large government or military software contracts, especially when their
funding comes from legislation. For such projects, funds have to be allocated before the work starts. Also, as said in the introduction,
for such large government  waterfall projects, it is  often required that the proposed budget is   double-checked by some estimation model~\cite{MenziesNeg:2017}.  For these reasons:
\begin{quote}
{\em Effort estimation methods need to support {\em both} classic waterfall projects
{\em and} contemporary projects.}
\end{quote}
We note that effort estimation in software development can be categorized into human-based and algorithm-based methods~\cite{teak2012,shepperd2007software}. In this paper we focus on algorithm-based methods since they are preferred when estimates have to be audited or debated (these methods are explicit and available for inspection). To understand the range of possible estimates, we can run the algorithm as many times as necessary, which may not be applicable by using human-based methods. Algorithm-based methods can have comparable performance to human-based ones. J{\o}rgensen et al. indicate that even very strong
advocates of human-based methods acknowledge that algorithm-based methods are useful for learning the uncertainty about particular estimates~\cite{jorgensen2009impact}.

Algorithm-based   methods have been widely explored in the past few decades including    classic  model like COCOMO and more recent  proposals   like ATLM~\cite{Whigham:2015} and LP4EE~\cite{SarroTOSEM2018}.

\subsection{COCOMO}
\label{sec:coco}

COCOMO (the COnstructive COst MOdel) is a procedural cost estimate model for software projects proposed by Boehm et al. based on LOC (number of Lines of Code). It is often used as a process of reliably predicting the various parameters associated with making a project such as size, effort, cost, time and quality. In late 1970s, Boehm was able to gather 63 project data points that could be published and to extend the model to
include alternative development modes that covered
other types of software such as business data
processing.  The resulting model was called the
Constructive Cost Model, or COCOMO, and was
published along with the data in the book Software
Engineering Economics~\cite{boehm1981software}. 
In this first version model (COCOMO-I), project attributes
were scored using just a few coarse-grained values (very low,
low, nominal, high, very high). These attributes
are {\em effort multipliers} where
a off-nominal value changes the estimate by some number
greater or smaller than one.
In COCOMO-I,  all attributes (except KLOC)
effect effort linearly.

Boehm created a consortium for
industrial organizations after COCOMO was released.
It collected information on 161 projects from commercial,
aerospace, government, and non-profit organizations.
Based on an analysis of those 161 projects, new attributes called {\em scale factors} were added to the original model, which had an {\em exponential impact}
on effort.
Using the new data, Boehm et al. developed COCOMO-II model that map the project descriptors (very low, low, etc.)
into the specific values~\cite{boehm2000cost}:
\begin{equation}\label{eq:coc}
\mathit{effort}=a\prod_i EM_i *\mathit{KLOC}^{b+0.01\sum_j SF_j}
\end{equation}
Inside this equation, $a,b$ are the {\em local calibration} parameters (with default values of 2.94 and 0.91). {\em EM} stands for effort multipliers, and {\em SF} are scale factors.  Boehm offers a simple linear time {\em local calibration} procedure~\cite{boehm2000cost} to update these defaults using the local training data.
The calculated {\em effort}
measures ``development months'' where one month is 152 hours of work  (and includes development and management hours). For details about COCOMO attributes, see tiny.cc/ccm\_attr.


\subsection{Beyond COCOMO}

For modern software development, it is necessary to develop new technique and make changes to improve COCOMO-style estimation. Robles et al. report that more  companies are turning to open source software projects (e.g. Contemporary software projects on Github), other than traditional waterfall style projects for their new business strategy~\cite{robles2014estimating}. For old parametric estimating models like COCOMO, Shepperd et al. found it is difficult to determine some of their features for the estimations~\cite{shepperd2007software}. COCOMO measured software size by using LOC (line of code), but this feature is not available during the coding procedure, and it is difficult to make comparisons between different programming languages that may take varying numbers of statements to perform a given function. Jeffery et al. indicated that parametric model like COCOMO need to be calibrated to be used effectively in their study~\cite{jeffery1990calibrating}, which is another evidence that old parametric estimating models like COCOMO may not be appropriate for newer tasks. 


\subsubsection{ATLM}
\label{sec:atlm}
Automatically Transformed Linear Model (ATLM) is a multiple linear regression model proposed by Whigham et al.~\cite{Whigham:2015}. It calculates the effort as:

\[
\mathit{effort} = \beta_0 + \sum_i\beta_i\times a_{i} +  \varepsilon_i
\]
where $a_i$ is explanatory attribute and $\varepsilon_i$ is error to the actual value. The prediction weight $\beta_i$ is determined using least square error estimation~\cite{neter1996applied}. Additionally, transformations are applied on the attributes to further minimize the error in the model. In case of categorical attributes, the standard approach of ``dummy variables"~\cite{hardy1993regression} is applied. While, for continuous attributes, transformations such as logarithmic, square root,  or no transformation is employed such that the skewness of the attribute is minimum. 

It should be noted that, ATLM does not consider relatively complex techniques like using model residuals,  box transformations or step-wise regression (which are standard) when developing a linear regression model. The authors make this decision since they intend ATLM to be a simple baseline model rather than the ``best" model.

\subsubsection{LP4EE}
\label{sec:lp4ee}
Linear Programming for Effort Estimation (LP4EE) is a newly developed method by Sarro et al.~\cite{SarroTOSEM2018}, it aims to achieve the best outcome from a mathematical
model with a linear objective function subject to linear equality and inequality
constraints. The feasible region is given by the intersection of the constraints and the
Simplex (linear programming algorithm) is able to find a point in the polyhedron where
the function has the smallest error in polynomial time. In effort estimation problem, this model minimizes the Sum of Absolute Residual (SAR), when a new project is presented to the model, LP4EE predicts the effort as

\[
\mathit{effort} = a_1*x_1 + a_2*x_2 + ... + a_n*x_n
\]
where $x_i$ is the value of a given project feature and $a_i$ is the corresponding coefficient evaluated by linear programming.
Sarro et al. propose LP4EE as another baseline model for effort estimation since it provides
similar or more accurate estimates than ATLM and is much less sensitive than ATLM
to multiple data splits and different cross-validation methods\cite{SarroTOSEM2018}.

\subsubsection{Machine Learning-based Effort Estimators}
\label{sec:algo}

\newcommand{\centered}[1]{\begin{tabular}{l} #1 \end{tabular}}

Many machine learning algorithms have been used for software effort estimation. 
Random Forest~\cite{breiman2001random} and Support Vector Regression~\cite{chang2011libsvm} are such instances of regression methods. Random Forest (RF) is an ensemble learning method for  regression (and classification) tasks that  builds a set of   trees when training the model. To make the final prediction , it uses 
the mode of the classes (classification) or mean prediction (regression) of the individual trees.
Support Vector Regression (SVR) uses kernel functions to project
the data onto a new hyperspace where complex non-linear patterns
can be simply represented.  
Another  learning approach is to use
a $K=5$ nearest-neighbor method~\cite{shepperd1997estimating}. 
For each test instance, KNN then selects   $k$ similar analogies out of a training set. The resultant prediction is the the mean of the 
class value of those $k$ neighbors.

Some algorithm-based estimators use regression trees such as CART~\cite{brieman84}.
CART is a  tree learner that divides a data set, then recurses
on each split.
If data contains more than {\em min\_sample\_split}, then a split is attempted.
On the other hand, if a split contains no more than {\em min\_samples\_leaf}, then the recursion stops. 
CART finds the attributes whose ranges contain rows with least variance in the number
of defects. If an  attribute ranges $r_i$ is found in 
$n_i$ rows each with an effort  variance of $v_i$, then CART seeks the  attribute with a split that most
minimizes $\sum_i \left(\sqrt{v_i}\times n_i/(\sum_i n_i)\right)$.
For more details on the CART parameters, see Table~\ref{tbl:cart}. Note that we choose the tuning range by using advice from Fu et al.~\cite{Fu2016TuningFS}.


\begin{table}[!htbp]
\caption{CART's parameters.}
\begin{adjustbox}{max width=0.48\textwidth}
\begin{tabular}{l|c|c|c|l}
\hline
\rowcolor[HTML]{EFEFEF} 
\multicolumn{1}{c|}{\cellcolor[HTML]{EFEFEF}\textbf{Parameter}} & \textbf{Type} & \textbf{Default} & \textbf{Tuning Range} & \multicolumn{1}{c}{\cellcolor[HTML]{EFEFEF}\textbf{Description}} \\ \hline
max\_feature & numerical & None & {[}0.01, 1{]} & \begin{tabular}[c]{@{}l@{}}Number of features to consider\\ when looking for the best split\end{tabular} \\ \hline
max\_depth & numerical & None & {[}1, 12{]} & \begin{tabular}[c]{@{}l@{}}The maximum depth of the\\ decision tree\end{tabular} \\ \hline
min\_sample\_split & numerical & 2 & {[}0, 20{]} & \begin{tabular}[c]{@{}l@{}}Minimum samples required to \\ split internal nodes\end{tabular} \\ \hline
min\_sample\_leaf & numerical & 1 & {[}1, 12{]} & \begin{tabular}[c]{@{}l@{}}Minimum samples required to\\ be at a leaf node\end{tabular} \\ \hline
\end{tabular}
\label{tbl:cart}
\end{adjustbox}
\end{table}

Before moving on from CART, we note a detail that will become important when we discuss our third research question.  Note that   {\em decreasing} max\_depth and {\em increasing} min\_sample\_leaf will result in smaller trees. In such smaller trees, few features will appear; specifically, on those features
that most minimize the standard deviation of the target class. 
 In the experimental rig described below, many times, we will generate trees using different settings to \tbl{cart}. By counting the the number of times a feature appears in these trees, we can infer what features are the most important to effort estimation.

\subsubsection{Hyperparameter Optimization} \label{sec:tuning}


Hyperparameters control the algorithm policies of the learners.  Choosing appropriate hyperparameters plays a critical role in the performance of machine learning models. Tuning hyperparameters is the process of searching the most optimal hyperparameter options for machine learning models~\cite{biedenkapp2018hyperparameter,franceschi2017forward}.
Some popular methods to tune the hyperparameters are grid search and differential evolution.

\textit{Grid search}~\cite{bergstra2011algorithms} is a technique that using brute force of all combinations for hyperparameters. Although the Grid search method is a simple algorithm to use, it suffers if data have high dimensional space called the ``curse of dimensionality''. Previous work has shown that grid search might also miss important optimizations~\cite{fu2016differential} or run needlessly slowly since,
often, only a few of the tuning parameters really matter~\cite{Bergstra:2012}.


\textit{Differential evolution} (DE)~\cite{storn1997differential}.
 The premise of DE is that the best way to mutate the existing tunings is to extrapolate between current solutions. Three solutions $a, b, c$ are selected at random. For each tuning parameter $k$, at some probability $cr$, we replace the old tuning $x_k$ with $y_k$ where
\mbox{$y_k = a_k + f \times (b_k - c_k)$}
where $f$ is a parameter controlling differential weight.  
The main loop of DE runs over the population of size $np$, replacing old items with new candidates (if new candidate is better). This means that, as the loop progresses, the population is full of increasingly more valuable solutions (which, in turn,
helps   extrapolation).

\textit{Bayesian optimization}~\cite{pelikan1999simple} works by assuming the unknown function was sampled from a Gaussian Process and maintains a posterior distribution for this function as observation are made. However, it might not be well-suited for optimization over continuous domains with large number of dimensions~\cite{frazier2018tutorial}.

\begin{figure}
\centerline{\includegraphics[width=0.5\textwidth]{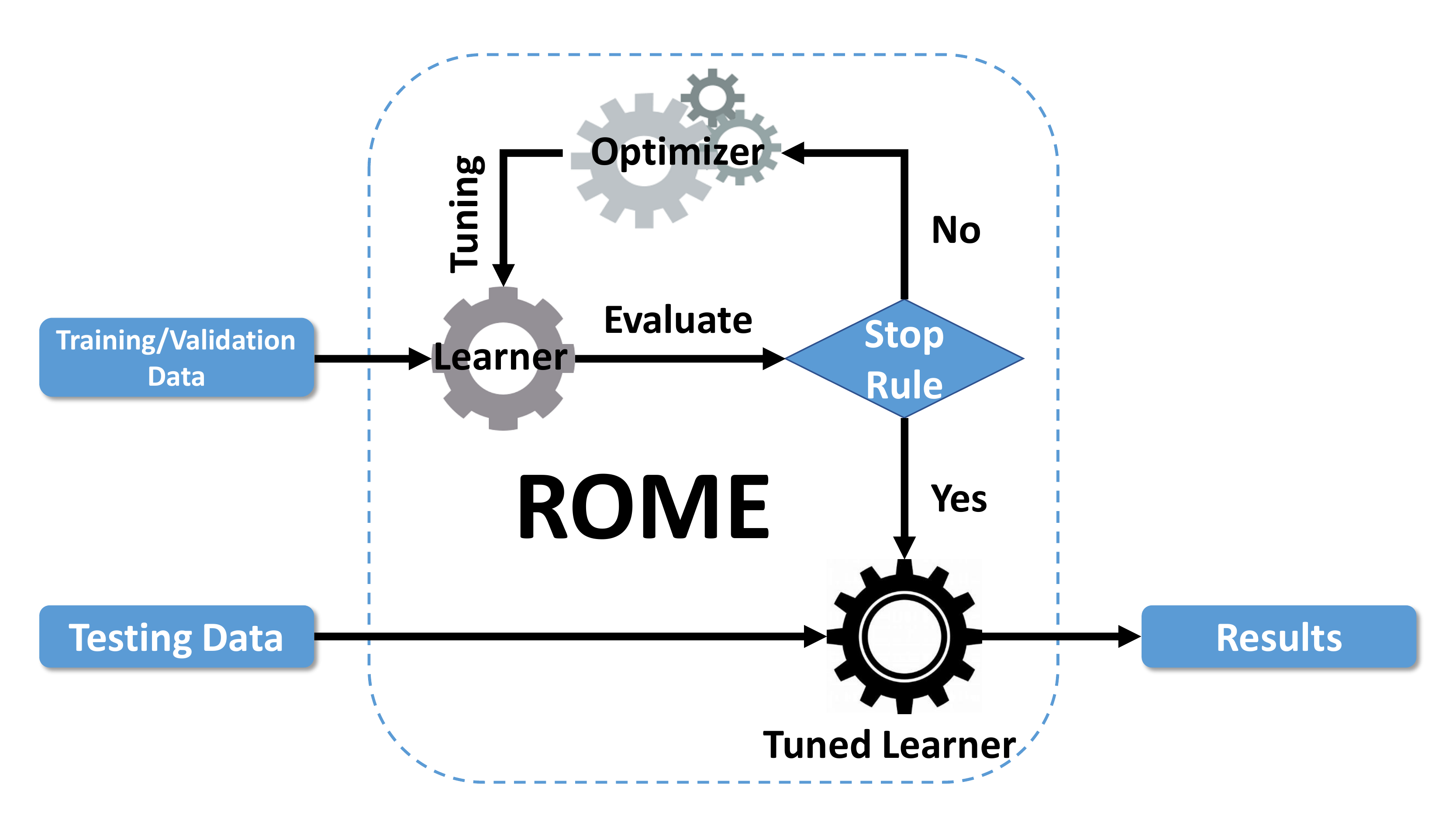}}
\caption{ROME's architecture}    
\label{fig:romearc}
\end{figure}

\subsubsection{ROME}
\label{sec:optim}

Standard hyperparameter optimization with DE or Bayesian optimization can be a tedious and time consuming task~\cite{fu2016differential}. 
Utilizing sequential model optimization (SMO), FLASH, introduced by Vivek et al.~\cite{nair2017flash}, terminates after just a few dozen executions of different learner
control parameters.  
ROME uses an adjusted version of FLASH as an optimizer to tune CART~\cite{brieman84}.

As shown in  \fig{romearc}, ROME has a   learning layer and a optimizing layer. When training data arrives, the estimator in the learning layer is being trained, and the optimizer in optimizing layer provides better hyperparameters to the learner to help improve the performance of estimators. Such trained learner will be evaluated on the validation data afterwards.  Once some stopping criteria is met, the generated learner is then passed to the test data for final testing.

When we design ROME, we want it to be as flexible as possible. It was simple  to ``pop the top'' and replace the optimizing layer with another optimizer. In this paper, ROME uses FLASH~\cite{nair2017flash} as the optimizer (with suitable parameter settings to adjust the data). Since the result from that initial study were promising, we paused further experimentation to record those results.
In future work, we will try other optimizers.


\begin{table*}
\centering
\caption{\textcolor{black}{ Some data from the  NASA10 data set (one row per project). For a definition of the terms in row1 (``prec'', ``flex'', ``resl'' etc.) see tiny.cc/ccm\_attr.
As to the different columns, scale factors change effort exponentially while effort multipliers have a linear impact on effort.
Any effort multiplier with a value of ``3'' is a {\em nominal} value; i.e. it multiplies the effort by a multiple of 1.0. Effort multipliers
above and below ``3'' can each effect project effort by a multiple ranging from 0.7 to 1.74.  
}}\label{tbl:nasa10}
    
  \footnotesize
  \setlength{\tabcolsep}{2.5pt}
\begin{tabu}{|ccccc|ccccccccccccccccc|c|c|}\hline
\rowfont{\color{white}} \rowcolor{black!70} prec&flex&resl&team&pmat&rely&cplx&data&ruse&     time&stor&pvol&acap&pcap&pcon&aexp&plex&     ltex&tool&sced&site&docu&kloc&months\\\hline
2&2&2&3&3&4&5&4&3&5&6&4&4&4&3&4&3&3&1&3&4&4&77&1830\\
\rowcolor{black!20}2&2&2&3&3&5&5&2&3&5&6&2&4&3&3&2&1&2&2&3&4&4&24&648\\
2&2&2&3&3&4&5&3&3&5&5&4&3&3&3&3&2&2&1&3&4&4&23&492\\
\rowcolor{black!20}2&2&3&3&2&4&4&3&2&3&3&4&3&3&3&3&3&4&2&3&5&3&146&3292\\
2&3&3&5&3&3&4&3&2&4&4&2&5&5&4&5&1&5&3&3&6&3&113&1080\\\hline
\noalign{\vspace{-7pt}}
\multicolumn{5}{c}{$\underbrace{\hspace*{28\tabcolsep}}$}&
\multicolumn{17}{c}{$\underbrace{\hspace*{120\tabcolsep}}$}&
\multicolumn{1}{c}{$\underbrace{\hspace*{5\tabcolsep}}$} &
\multicolumn{1}{c}{$\underbrace{\hspace*{5\tabcolsep}}$}\\
\multicolumn{5}{c}{\bf \normalsize scale factors} & \multicolumn{17}{c}{\bf \normalsize effort multipliers} & \multicolumn{1}{c}{\bf \normalsize size} & \multicolumn{1}{c}{\bf \normalsize effort}\\
\end{tabu}
\end{table*}

\begin{table*} 
\centering
\caption{Descriptive Statistics of the classic effort data sets. Terms in \IT{italic} are removed from this study, for reasons discussed in the text.}\label{table:classic}
\renewcommand{\baselinestretch}{0.75} 
\resizebox{0.95\textwidth}{!}{
\centering
\begin{tabular}{cc}
\scriptsize
\begin{tabular}{|c|l|rrrr|}
    \hline
      & feature & min  & max & mean & std\\
   \hline


\multirow{7}{*}{\begin{sideways}kemerer\end{sideways}}
& Langu. & 1 & 3 & 1.2 & 0.6\\
& Hdware & 1 & 6 & 2.3 & 1.7\\
& \IT{Duration} & 5 & 31 & 14.3 & 7.5\\
& \IT{KSLOC} & 39 & 450 & 186.6 & 136.8\\
& AdjFP & 100 & 2307 & 999.1 & 589.6\\
& \IT{RAWFP} & 97 & 2284 & 993.9 & 597.4\\
& Effort & 23 & 1107 & 219.2 & 263.1\\
\hline
\multirow{8}{*}{\begin{sideways}albrecht\end{sideways}}
& Input & 7 & 193 & 40.2 & 36.9\\
& Output & 12 & 150 & 47.2 & 35.2\\
& Inquiry & 0 & 75 & 16.9 & 19.3\\
& File & 3 & 60 & 17.4 & 15.5\\
& \IT{FPAdj} & 1 & 1 & 1.0 & 0.1\\
& \IT{RawFPs} & 190 & 1902 & 638.5 & 452.7\\
& \IT{AdjFP} & 199 & 1902 & 647.6 & 488.0\\
& Effort & 0 & 105 & 21.9 & 28.4\\
\hline
\multirow{12}{*}{\begin{sideways}isbsg10\end{sideways}}
& UFP & 1 & 2 & 1.2 & 0.4\\
& IS & 1 & 10 & 3.2 & 3.0\\
& DP & 1 & 5 & 2.6 & 1.1\\
& LT & 1 & 3 & 1.6 & 0.8\\
& PPL & 1 & 14 & 5.1 & 4.1\\
& CA & 1 & 2 & 1.1 & 0.3\\
& FS & 44 & 1371 & 343.8 & 304.2\\
& RS & 1 & 4 & 1.7 & 0.9\\
& FPS & 1 & 5 & 3.5 & 0.7\\
& Effort & 87 & 14453 & 2959 & 3518\\
\hline
\multirow{8}{*}{\begin{sideways}finnish\end{sideways}}
& hw & 1 & 3 & 1.3 & 0.6\\
& at & 1 & 5 & 2.2 & 1.5\\
& FP & 65 & 1814 & 763.6 & 510.8\\
& co & 2 & 10 & 6.3 & 2.7\\
& \IT{prod} & 1 & 29 & 10.1 & 7.1\\
& \IT{lnsize} & 4 & 8 & 6.4 & 0.8\\
& \IT{lneff} & 6 & 10 & 8.4 & 1.2\\
& Effort & 460 & 26670 & 7678 & 7135\\
\hline

\end{tabular} 


~

\scriptsize
\begin{tabular}{|c|l|rrrr|}
    \hline
      & feature
    & min  & max & mean & std\\
   \hline

\multirow{8}{*}{\begin{sideways}miyazaki\end{sideways}}
& \IT{KLOC} & 7 & 390 & 63.4 & 71.9\\
& SCRN & 0 & 150 & 28.4 & 30.4\\
& FORM & 0 & 76 & 20.9 & 18.1\\
& FILE & 2 & 100 & 27.7 & 20.4\\
& ESCRN & 0 & 2113 & 473.0 & 514.3\\
& EFORM & 0 & 1566 & 447.1 & 389.6\\
& EFILE & 57 & 3800 & 936.6 & 709.4\\
& Effort & 6 & 340 & 55.6 & 60.1\\
\hline
\multirow{26}{*}{\begin{sideways}maxwell\end{sideways}}
& App & 1 & 5 & 2.4 & 1.0\\
& Har & 1 & 5 & 2.6 & 1.0\\
& Dba & 0 & 4 & 1.0 & 0.4\\
& Ifc & 1 & 2 & 1.9 & 0.2\\
& Source & 1 & 2 & 1.9 & 0.3\\
& Telon. & 0 & 1 & 0.2 & 0.4\\
& Nlan & 1 & 4 & 2.5 & 1.0\\
& T01 & 1 & 5 & 3.0 & 1.0\\
& T02 & 1 & 5 & 3.0 & 0.7\\
& T03 & 2 & 5 & 3.0 & 0.9\\
& T04 & 2 & 5 & 3.2 & 0.7\\
& T05 & 1 & 5 & 3.0 & 0.7\\
& T06 & 1 & 4 & 2.9 & 0.7\\
& T07 & 1 & 5 & 3.2 & 0.9\\
& T08 & 2 & 5 & 3.8 & 1.0\\
& T09 & 2 & 5 & 4.1 & 0.7\\
& T10 & 2 & 5 & 3.6 & 0.9\\
& T11 & 2 & 5 & 3.4 & 1.0\\
& T12 & 2 & 5 & 3.8 & 0.7\\
& T13 & 1 & 5 & 3.1 & 1.0\\
& T14 & 1 & 5 & 3.3 & 1.0\\
& \IT{Dura.} & 4 & 54 & 17.2 & 10.7\\
& Size & 48 & 3643 & 673.3 & 784.1\\
& \IT{Time} & 1 & 9 & 5.6 & 2.1\\
& Effort & 583 & 63694 & 8223 & 10500\\
\hline

\end{tabular}


~

\scriptsize
\begin{tabular}{|c|l|rrrr|}
    \hline
      & feature
    & min  & max & mean & std\\
  \hline

\multirow{7}{*}{\begin{sideways}desharnais\end{sideways}}
& TeamExp & 0 & 4 & 2.3 & 1.3\\
& MngExp & 0 & 7 & 2.6 & 1.5\\
& \IT{Length} & 1 & 36 & 11.3 & 6.8\\
& Trans.s & 9 & 886 & 177.5 & 146.1\\
& Entities & 7 & 387 & 120.5 & 86.1\\
& AdjPts & 73 & 1127 & 298.0 & 182.3\\
& Effort & 546 & 23940 & 4834 & 4188\\
\hline
\multirow{7}{*}{\begin{sideways}kitchenham\end{sideways}}
& code & 1 & 6 & 2.1 & 0.9\\
& type & 0 & 6 & 2.4 & 0.9\\
& \IT{duration} & 37 & 946 & 206.4 & 134.1\\
& fun\_pts & 15 & 18137 & 527.7 & 1522\\
& \IT{estimate} & 121 & 79870 & 2856 & 6789\\
& \IT{esti\_mtd} & 1 & 5 & 2.5 & 0.9\\
& Effort & 219 & 113930 & 3113 & 9598\\
\hline
\multirow{19}{*}{\begin{sideways}china\end{sideways}}
& \IT{ID} & 1 & 499 & 250.0 & 144.2\\
& \IT{AFP} & 9 & 17518 & 486.9 & 1059\\
& Input & 0 & 9404 & 167.1 & 486.3\\
& Output & 0 & 2455 & 113.6 & 221.3\\
& Enquiry & 0 & 952 & 61.6 & 105.4\\
& File & 0 & 2955 & 91.2 & 210.3\\
& Interface & 0 & 1572 & 24.2 & 85.0\\
& \IT{Added} & 0 & 13580 & 360.4 & 829.8\\
& \IT{changed} & 0 & 5193 & 85.1 & 290.9\\
& \IT{Deleted} & 0 & 2657 & 12.4 & 124.2\\
& \IT{PDR\_A} & 0 & 84 & 11.8 & 12.1\\
& \IT{PDR\_U} & 0 & 97 & 12.1 & 12.8\\
& \IT{NPDR\_A} & 0 & 101 & 13.3 & 14.0\\
& \IT{NPDU\_U} & 0 & 108 & 13.6 & 14.8\\
& Resource & 1 & 4 & 1.5 & 0.8\\
& \IT{Dev.Type} & 0 & 0 & 0.0 & 0.0\\
& \IT{Duration} & 1 & 84 & 8.7 & 7.3\\
& \IT{N\_effort} & 31 & 54620 & 4278 & 7071\\
& Effort & 26 & 54620 & 3921 & 6481\\
\hline

\end{tabular}


\end{tabular}
}
\end{table*}

FLASH comes from research into software configuration.   One of the new insights that leads to this paper was that ``configuration'' is a synonym for ``hyperparameter optimziation''. Hence, hyperparameter-optimization-via-configuration tools
has not previously been explored in the literature.  
Also, prior to this paper, such optimizers have not been used
for   effort estimation.

FLASH is a sequential model-based optimizer (SMO)~\cite{bergstra2011algorithms} (also known
in the machine learning literature
as an {\em active learner}~\cite{das16} or, in  the statistics literature  as 
{\em optimal experimental design}~\cite{olsson2009literature}). No matter whatever the name is, the idea behind it is the same:
reflect on the model built so far    to find the next best example
to evaluate. To tune a learning algorithm, FLASH explores $N$ possible tunings as follows:
\begin{enumerate}
\item
Set the evaluation budget $b$. Based on prior work~\cite{nair2017flash},   we used $b=200$.
 \item
Run the learning algorithm with $n=20$ to randomly select tunings.
\item Build an {\em archive} of  $n$   examples holding pairs of  parameter settings and   their resulting performance scores.
\item
Using that archive, learn a {\em surrogate}   to predicts performance. 
As per the methods of Nair et al.~\cite{nair2017flash}, our surrogates come from     CART~\cite{brieman84}.
\item Use the surrogate to guess  $M$ performance scores where
$M<N$ and $M \gg n$ parameter settings. Note that this step is very fast because all required is to run $M$ vectors downwards some very small CART trees.
\item use a {\em selection  function} to select  the most ``interesting'' setting.
We use the setting whose prediction has the smallest predicted error.
\item Collect performance scores by evaluating    ``interesting'' using
the data miners. Set $b=b-1$.
\item  Add  ``interesting'' to  archive. If  $b>0$, goto step 4.
\item
Else, halt.
\end{enumerate}


In summary,  given what we already know about the tunings (represented in a CART tree),
FLASH finds the potentially best tunings (in Step 6); then evaluate the performance~(in Step 7); 
then update the model with the results of that evaluation.

\section{Empirical Study} \label{sect:study} 

\begin{table*} 
\centering
\caption{Feature description of the Github data sets.}\label{table:dataset}

\renewcommand{\baselinestretch}{0.75} 
\resizebox{0.75\textwidth}{!}{
\centering
\begin{tabular}{|l|l|}
\hline
Feature & Description \\ \hline
dates & The end date of monthly data collection \\
monthly\_commits & Total number of commits created in last month \\
monthly\_commit\_comments & Total number of commit comments created in last month \\
monthly\_contributors & Total number of contributors that at least have one commit in last month \\
monthly\_open\_PRs & Total number of pull requests opened in last month \\
monthly\_closed\_PRs & Total number of pull requests closed in last month \\
monthly\_merged\_PRs & Total number of pull requests merged in last month \\
monthly\_PR\_mergers & Total number of pull requests mergers in last month \\
monthly\_PR\_comments & Total number of pull request comments created in last month \\
monthly\_open\_issues & Total number of issues opened in last month \\
monthly\_closed\_issues & Total number of issues closed in last month \\
monthly\_issue\_comments & Total number of issue comments created in last month \\
monthly\_stargazer & Total number of new stars acquired in last month \\
monthly\_forks & Total number of forks occurred in last month \\
monthly\_watchers & Total number of new watchers acquired in last month \\ \hline
\end{tabular}
}
\end{table*}

\subsection{Data}

To evaluate the proposed ROME framework comprehensively, we test it out on both COCOMO-style data and non COCOMO-style data. For COCOMO-style data, we include 216 projects from the SEACRAFT repository\footnote{http://tiny.cc/seacrafts}; In \tbl{nasa10}, we list a sample of our data. This data set has been widely used to evaluate effort estimation methods for COCOMO-sytle data, which serves the same purpose to compare our proposed framework with the COCOMO-II procedure. 

To test how ROME performs on non COCOMO data, we use 945 classic effort projects from the SEACRAFT (described in Table~\ref{table:classic}), plus data collected from 120 repositories on Github by using Github API v3 \& GraphQL API v4. We collect 15 features from these Github repositories, for feature details, See Table~\ref{table:dataset}.

Note that some features of these non COCOMO style data sets are not used in our experiment because they are (1) naturally irrelevant to their effort values (e.g., ID, Syear), (2) unavailable at the prediction phase (e.g., duration, LOC), (3) highly correlated or overlap to each other (e.g., raw function point and adjusted function points). A data cleaning process is applied to solve this issue. 
Those removed features are highlighted as italic in Table~\ref{table:classic}. 

For contemporary data sets, we use number of monthly commits to measure developing effort of the repositories. Github uses commits to indicate change made by contributors to the source code of repositories, unlike some features like ``LOC'', commits are independent of programming languages, it has been used to measure the productivity of software developers in previous research~\cite{vasilescu2016sky}. In our experiments, we pick a random time point in the developing process of repositories, and predict the effort in the next month-long period. That is:
\bi
\item When estimating contemporary Github projects, we use the last three months of data
to predict the effort in the next month;
\item On the other hand, when estimating classic waterfall projects, we build an estimator
for completing the entire task (using the methods described in the next section).
\ei

\subsection{Experimental Rig}

In our experiments related to COCOMO and SEACRAFT data set, we use a {\em M*N-way} cross-validation to split training and testing data for the estimators. That is, in $M$ times,  shuffle the data randomly (using a different random number seed)
then divide the data into $N$ bins.
For $i   \in N$, bin $i$ is used to test a model
build from the other bins.
Following the advice
of Nair et al.~\cite{nair18}, we  use $N=3$  and $M=20$ for these effort data sets. 

For data sets collected from Github repositories, since they are time-series data, applying cross-validation is not appropriate. Thus, for each repository, we use last month's data as testing data and rest as training data. 

As a procedural detail, first we divided the data and then we applied the treatments. That is, all treatments saw the same training and test data.




In this experiment,  we do not tune    ATLM or LP4EE since they were designed to be used ``off-the-shelf'' (Whigham et al.~\cite{Whigham:2015} declare that one of ATLM's most important features is that it does not need tuning).
We also do not tune SVR and RF since we treat them as baseline algorithm-based methods in our benchmarks (i.e. use default settings in scikit-learn for these algorithms). Here, we add KNN and CART with default settings, since these methods often appear in effort estimation literature~\cite{MenziesNeg:2017,SarroTOSEM2018,teak2012,sarro2016multi}.
As to COCOMO-II, we apply Boehm's local calibration procedure~\cite{boehm2000cost} on the training data to  adjust the $(a,b)$ parameters of \eq{coc}.
Lastly, we compare the performance of ROME (CART with optimizer FLASH), to that of Differential Evolution~\cite{storn1997differential}.
 Using advice from Storn and Fu et al.~\cite{storn1997differential,Fu2016TuningFS}, for DE we use   $\{\mathit{np,g,cr,\mathit{generations}}\}=\{20,0.75,0.3,10\}$.

\begin{table*} 
\centering
\caption{MRE (Magnitude of the Relative Error) of classic waterfall data sets, \underline{lower} values are \underline{better}. For each row, the gray cells show the results that are statistically significantly  better than others on that row (as judged by a Scott-Knot bootstrap test plus an  A12 effect size test).
If multiple treatments tied for ``best'', then there will be multiple gray cells in a row.  }
\label{table:result_mre}
\begin{adjustbox}{max width=0.95\textwidth}
\centering
\begin{tabular}{clllllllllllllllllll}
\hline
\multicolumn{2}{|c|}{\cellcolor[HTML]{EFEFEF}} & \multicolumn{8}{c|}{\cellcolor[HTML]{EFEFEF}\textbf{Scikit-Learn}} & \multicolumn{4}{c|}{\cellcolor[HTML]{EFEFEF}\textbf{Tuned}} & \multicolumn{4}{c|}{\cellcolor[HTML]{EFEFEF}\textbf{Other Methods}} & \multicolumn{2}{c|}{\cellcolor[HTML]{EFEFEF}\textbf{ COCOMO}} \\ \cline{3-20} 
\multicolumn{2}{|c|}{\multirow{-2}{*}{\cellcolor[HTML]{EFEFEF}\textbf{Dataset}}} & \multicolumn{2}{c|}{\textbf{KNN}} & \multicolumn{2}{c|}{\textbf{SVR}} & \multicolumn{2}{c|}{\textbf{CART}} & \multicolumn{2}{c|}{\textbf{RF}} & \multicolumn{2}{c|}{\textbf{CART\_DE}} & \multicolumn{2}{c|}{\textbf{ROME} } & \multicolumn{2}{c|}{\textbf{ATLM}} & \multicolumn{2}{c|}{\textbf{LP4EE}} & \multicolumn{2}{c|}{\textbf{COCOMO-II}} \\ \hline
\multicolumn{1}{|c|}{\cellcolor[HTML]{EFEFEF}} & \multicolumn{1}{l|}{kemerer} & \multicolumn{2}{c|}{0.56} & \multicolumn{2}{c|}{0.59} & \multicolumn{2}{c|}{0.55} & \multicolumn{2}{c|}{0.50} & \multicolumn{2}{c|}{\cellcolor[HTML]{C0C0C0}0.32} & \multicolumn{2}{c|}{\cellcolor[HTML]{C0C0C0}0.37} & \multicolumn{2}{c|}{0.76} & \multicolumn{2}{c|}{0.54} & \multicolumn{2}{c|}{N/A} \\
\multicolumn{1}{|c|}{\cellcolor[HTML]{EFEFEF}} & \multicolumn{1}{l|}{albrecht} & \multicolumn{2}{c|}{0.45} & \multicolumn{2}{c|}{0.56} & \multicolumn{2}{c|}{0.53} & \multicolumn{2}{c|}{0.46} & \multicolumn{2}{c|}{\cellcolor[HTML]{C0C0C0}{\color[HTML]{333333} 0.32}} & \multicolumn{2}{c|}{\cellcolor[HTML]{C0C0C0}0.33} & \multicolumn{2}{c|}{1.40} & \multicolumn{2}{c|}{0.44} & \multicolumn{2}{c|}{N/A} \\
\multicolumn{1}{|c|}{\cellcolor[HTML]{EFEFEF}} & \multicolumn{1}{l|}{isbsg10} & \multicolumn{2}{c|}{0.73} & \multicolumn{2}{c|}{0.72} & \multicolumn{2}{c|}{0.74} & \multicolumn{2}{c|}{0.78} & \multicolumn{2}{c|}{\cellcolor[HTML]{C0C0C0}0.59} & \multicolumn{2}{c|}{\cellcolor[HTML]{C0C0C0}0.62} & \multicolumn{2}{c|}{1.27} & \multicolumn{2}{c|}{0.75} & \multicolumn{2}{c|}{N/A} \\
\multicolumn{1}{|c|}{\cellcolor[HTML]{EFEFEF}} & \multicolumn{1}{l|}{finnish} & \multicolumn{2}{c|}{0.64} & \multicolumn{2}{c|}{0.74} & \multicolumn{2}{c|}{0.57} & \multicolumn{2}{c|}{0.57} & \multicolumn{2}{c|}{0.48} & \multicolumn{2}{c|}{\cellcolor[HTML]{C0C0C0}0.42} & \multicolumn{2}{c|}{0.87} & \multicolumn{2}{c|}{0.63} & \multicolumn{2}{c|}{N/A} \\
\multicolumn{1}{|c|}{\cellcolor[HTML]{EFEFEF}} & \multicolumn{1}{l|}{miyazaki} & \multicolumn{2}{c|}{0.47} & \multicolumn{2}{c|}{0.37} & \multicolumn{2}{c|}{0.47} & \multicolumn{2}{c|}{0.46} & \multicolumn{2}{c|}{\cellcolor[HTML]{C0C0C0}0.32} & \multicolumn{2}{c|}{\cellcolor[HTML]{C0C0C0}0.32} & \multicolumn{2}{c|}{0.37} & \multicolumn{2}{c|}{\cellcolor[HTML]{C0C0C0}0.33} & \multicolumn{2}{c|}{N/A} \\
\multicolumn{1}{|c|}{\cellcolor[HTML]{EFEFEF}} & \multicolumn{1}{l|}{maxwell} & \multicolumn{2}{c|}{0.56} & \multicolumn{2}{c|}{0.56} & \multicolumn{2}{c|}{0.52} & \multicolumn{2}{c|}{0.51} & \multicolumn{2}{c|}{\cellcolor[HTML]{C0C0C0}0.38} & \multicolumn{2}{c|}{\cellcolor[HTML]{C0C0C0}0.36} & \multicolumn{2}{c|}{2.82} & \multicolumn{2}{c|}{0.51} & \multicolumn{2}{c|}{N/A} \\
\multicolumn{1}{|c|}{\cellcolor[HTML]{EFEFEF}} & \multicolumn{1}{l|}{desharnais} & \multicolumn{2}{c|}{0.50} & \multicolumn{2}{c|}{0.48} & \multicolumn{2}{c|}{0.49} & \multicolumn{2}{c|}{0.46} & \multicolumn{2}{c|}{\cellcolor[HTML]{C0C0C0}0.35} & \multicolumn{2}{c|}{\cellcolor[HTML]{C0C0C0}0.35} & \multicolumn{2}{c|}{0.54} & \multicolumn{2}{c|}{\cellcolor[HTML]{C0C0C0}0.38} & \multicolumn{2}{c|}{N/A} \\
\multicolumn{1}{|c|}{\cellcolor[HTML]{EFEFEF}} & \multicolumn{1}{l|}{kitchenham} & \multicolumn{2}{c|}{0.39} & \multicolumn{2}{c|}{0.60} & \multicolumn{2}{c|}{0.49} & \multicolumn{2}{c|}{0.43} & \multicolumn{2}{c|}{0.38} & \multicolumn{2}{c|}{\cellcolor[HTML]{C0C0C0}0.34} & \multicolumn{2}{c|}{1.06} & \multicolumn{2}{c|}{0.38} & \multicolumn{2}{c|}{N/A} \\
\multicolumn{1}{|c|}{\multirow{-9}{*}{\cellcolor[HTML]{EFEFEF}\textbf{classic}}} & \multicolumn{1}{l|}{china} & \multicolumn{2}{c|}{0.64} & \multicolumn{2}{c|}{0.71} & \multicolumn{2}{c|}{0.71} & \multicolumn{2}{c|}{0.69} & \multicolumn{2}{c|}{0.64} & \multicolumn{2}{c|}{0.61} & \multicolumn{2}{c|}{\cellcolor[HTML]{C0C0C0}0.48} & \multicolumn{2}{c|}{\cellcolor[HTML]{C0C0C0}0.45} & \multicolumn{2}{c|}{N/A} \\
\multicolumn{1}{l}{} &  & \multicolumn{2}{l}{} & \multicolumn{2}{l}{} & \multicolumn{2}{l}{} & \multicolumn{2}{l}{} & \multicolumn{2}{l}{} & \multicolumn{2}{l}{} & \multicolumn{2}{l}{} & \multicolumn{2}{l}{} & \multicolumn{2}{l}{} \\ \hline
\multicolumn{1}{|c|}{\cellcolor[HTML]{EFEFEF}} & \multicolumn{1}{l|}{cocomo10} & \multicolumn{2}{c|}{0.67} & \multicolumn{2}{c|}{0.86} & \multicolumn{2}{c|}{0.33} & \multicolumn{2}{c|}{0.30} & \multicolumn{2}{c|}{0.30} & \multicolumn{2}{c|}{\cellcolor[HTML]{C0C0C0}0.28} & \multicolumn{2}{c|}{2.49} & \multicolumn{2}{c|}{0.32} & \multicolumn{2}{c|}{0.60} \\
\multicolumn{1}{|c|}{\cellcolor[HTML]{EFEFEF}} & \multicolumn{1}{l|}{cocomo81} & \multicolumn{2}{c|}{0.93} & \multicolumn{2}{c|}{0.89} & \multicolumn{2}{c|}{0.77} & \multicolumn{2}{c|}{0.76} & \multicolumn{2}{c|}{0.65} & \multicolumn{2}{c|}{0.64} & \multicolumn{2}{c|}{3.37} & \multicolumn{2}{c|}{0.65} & \multicolumn{2}{c|}{\cellcolor[HTML]{C0C0C0}0.49} \\
\multicolumn{1}{|c|}{\multirow{-3}{*}{\cellcolor[HTML]{EFEFEF}\textbf{cocomo}}} & \multicolumn{1}{l|}{nasa93} & \multicolumn{2}{c|}{0.70} & \multicolumn{2}{c|}{0.84} & \multicolumn{2}{c|}{\cellcolor[HTML]{C0C0C0}0.42} & \multicolumn{2}{c|}{\cellcolor[HTML]{C0C0C0}0.41} & \multicolumn{2}{c|}{\cellcolor[HTML]{C0C0C0}0.42} & \multicolumn{2}{c|}{\cellcolor[HTML]{C0C0C0}0.40} & \multicolumn{2}{c|}{0.90} & \multicolumn{2}{c|}{\cellcolor[HTML]{C0C0C0}0.38} & \multicolumn{2}{c|}{0.61} 
\end{tabular}

\end{adjustbox}
\end{table*}

\begin{table*}
\centering
\caption{SA (Standard Accuracy) of classic waterfall data sets, \underline{higher} values are \underline{better}. Same format as Table~\ref{table:result_mre}. }
\label{table:result_sa}
\begin{adjustbox}{max width=0.95\textwidth}
\centering
\begin{tabular}{clllllllllllllllllll}
\hline
\multicolumn{2}{|c|}{\cellcolor[HTML]{EFEFEF}} & \multicolumn{8}{c|}{\cellcolor[HTML]{EFEFEF}\textbf{Scikit-Learn}} & \multicolumn{4}{c|}{\cellcolor[HTML]{EFEFEF}\textbf{Tuned}} & \multicolumn{4}{c|}{\cellcolor[HTML]{EFEFEF}\textbf{Other Methods}} & \multicolumn{2}{c|}{\cellcolor[HTML]{EFEFEF}\textbf{COCOMO}} \\ \cline{3-20} 
\multicolumn{2}{|c|}{\multirow{-2}{*}{\cellcolor[HTML]{EFEFEF}\textbf{Dataset}}} & \multicolumn{2}{c|}{\textbf{KNN}} & \multicolumn{2}{c|}{\textbf{SVR}} & \multicolumn{2}{c|}{\textbf{CART}} & \multicolumn{2}{c|}{\textbf{RF}} & \multicolumn{2}{c|}{\textbf{CART\_DE}} & \multicolumn{2}{c|}{\textbf{ROME}} & \multicolumn{2}{c|}{\textbf{ATLM}} & \multicolumn{2}{c|}{\textbf{LP4EE}} & \multicolumn{2}{c|}{\textbf{COCOMO-II}} \\ \hline
\multicolumn{1}{|c|}{\cellcolor[HTML]{EFEFEF}} & \multicolumn{1}{l|}{kemerer} & \multicolumn{2}{c|}{0.38} & \multicolumn{2}{c|}{0.28} & \multicolumn{2}{c|}{0.42} & \multicolumn{2}{c|}{0.41} & \multicolumn{2}{c|}{\cellcolor[HTML]{C0C0C0}0.55} & \multicolumn{2}{c|}{0.43} & \multicolumn{2}{c|}{0.30} & \multicolumn{2}{c|}{0.40} & \multicolumn{2}{c|}{N/A} \\
\multicolumn{1}{|c|}{\cellcolor[HTML]{EFEFEF}} & \multicolumn{1}{l|}{albrecht} & \multicolumn{2}{c|}{0.51} & \multicolumn{2}{c|}{0.30} & \multicolumn{2}{c|}{0.41} & \multicolumn{2}{c|}{0.49} & \multicolumn{2}{c|}{\cellcolor[HTML]{C0C0C0}0.59} & \multicolumn{2}{c|}{\cellcolor[HTML]{C0C0C0}0.65} & \multicolumn{2}{c|}{0.34} & \multicolumn{2}{c|}{0.47} & \multicolumn{2}{c|}{N/A} \\
\multicolumn{1}{|c|}{\cellcolor[HTML]{EFEFEF}} & \multicolumn{1}{l|}{isbsg10} & \multicolumn{2}{c|}{0.28} & \multicolumn{2}{c|}{0.25} & \multicolumn{2}{c|}{0.20} & \multicolumn{2}{c|}{0.22} & \multicolumn{2}{c|}{\cellcolor[HTML]{C0C0C0}0.33} & \multicolumn{2}{c|}{\cellcolor[HTML]{C0C0C0}0.30} & \multicolumn{2}{c|}{\cellcolor[HTML]{C0C0C0}0.30} & \multicolumn{2}{c|}{0.22} & \multicolumn{2}{c|}{N/A} \\
\multicolumn{1}{|c|}{\cellcolor[HTML]{EFEFEF}} & \multicolumn{1}{l|}{finnish} & \multicolumn{2}{c|}{0.40} & \multicolumn{2}{c|}{0.24} & \multicolumn{2}{c|}{0.42} & \multicolumn{2}{c|}{0.44} & \multicolumn{2}{c|}{0.49} & \multicolumn{2}{c|}{\cellcolor[HTML]{C0C0C0}0.54} & \multicolumn{2}{c|}{0.41} & \multicolumn{2}{c|}{0.39} & \multicolumn{2}{c|}{N/A} \\
\multicolumn{1}{|c|}{\cellcolor[HTML]{EFEFEF}} & \multicolumn{1}{l|}{miyazaki} & \multicolumn{2}{c|}{0.45} & \multicolumn{2}{c|}{0.41} & \multicolumn{2}{c|}{0.41} & \multicolumn{2}{c|}{0.46} & \multicolumn{2}{c|}{\cellcolor[HTML]{C0C0C0}0.53} & \multicolumn{2}{c|}{\cellcolor[HTML]{C0C0C0}0.53} & \multicolumn{2}{c|}{\cellcolor[HTML]{C0C0C0}0.50} & \multicolumn{2}{c|}{\cellcolor[HTML]{C0C0C0}0.52} & \multicolumn{2}{c|}{N/A} \\
\multicolumn{1}{|c|}{\cellcolor[HTML]{EFEFEF}} & \multicolumn{1}{l|}{maxwell} & \multicolumn{2}{c|}{0.39} & \multicolumn{2}{c|}{0.30} & \multicolumn{2}{c|}{0.37} & \multicolumn{2}{c|}{0.44} & \multicolumn{2}{c|}{\cellcolor[HTML]{C0C0C0}0.51} & \multicolumn{2}{c|}{\cellcolor[HTML]{C0C0C0}0.55} & \multicolumn{2}{c|}{-1.07} & \multicolumn{2}{c|}{\cellcolor[HTML]{C0C0C0}0.52} & \multicolumn{2}{c|}{N/A} \\
\multicolumn{1}{|c|}{\cellcolor[HTML]{EFEFEF}} & \multicolumn{1}{l|}{desharnais} & \multicolumn{2}{c|}{0.44} & \multicolumn{2}{c|}{0.43} & \multicolumn{2}{c|}{0.39} & \multicolumn{2}{c|}{0.46} & \multicolumn{2}{c|}{\cellcolor[HTML]{C0C0C0}0.53} & \multicolumn{2}{c|}{\cellcolor[HTML]{C0C0C0}0.53} & \multicolumn{2}{c|}{0.37} & \multicolumn{2}{c|}{0.48} & \multicolumn{2}{c|}{N/A} \\
\multicolumn{1}{|c|}{\cellcolor[HTML]{EFEFEF}} & \multicolumn{1}{l|}{kitchenham} & \multicolumn{2}{c|}{\cellcolor[HTML]{C0C0C0}0.47} & \multicolumn{2}{c|}{0.32} & \multicolumn{2}{c|}{0.34} & \multicolumn{2}{c|}{0.41} & \multicolumn{2}{c|}{0.40} & \multicolumn{2}{c|}{\cellcolor[HTML]{C0C0C0}0.44} & \multicolumn{2}{c|}{-0.03} & \multicolumn{2}{c|}{\cellcolor[HTML]{C0C0C0}0.52} & \multicolumn{2}{c|}{N/A} \\
\multicolumn{1}{|c|}{\multirow{-9}{*}{\cellcolor[HTML]{EFEFEF}\textbf{classic}}} & \multicolumn{1}{l|}{china} & \multicolumn{2}{c|}{\cellcolor[HTML]{C0C0C0}0.28} & \multicolumn{2}{c|}{0.21} & \multicolumn{2}{c|}{0.12} & \multicolumn{2}{c|}{0.21} & \multicolumn{2}{c|}{\cellcolor[HTML]{C0C0C0}0.27} & \multicolumn{2}{c|}{\cellcolor[HTML]{C0C0C0}0.30} & \multicolumn{2}{c|}{0.12} & \multicolumn{2}{c|}{\cellcolor[HTML]{C0C0C0}0.32} & \multicolumn{2}{c|}{N/A} \\
\multicolumn{1}{l}{} &  & \multicolumn{2}{l}{} & \multicolumn{2}{l}{} & \multicolumn{2}{l}{} & \multicolumn{2}{l}{} & \multicolumn{2}{l}{} & \multicolumn{2}{l}{} & \multicolumn{2}{l}{} & \multicolumn{2}{l}{} & \multicolumn{2}{l}{} \\ \hline
\multicolumn{1}{|c|}{\cellcolor[HTML]{EFEFEF}} & \multicolumn{1}{l|}{cocomo10} & \multicolumn{2}{c|}{0.22} & \multicolumn{2}{c|}{0.14} & \multicolumn{2}{c|}{0.52} & \multicolumn{2}{c|}{\cellcolor[HTML]{C0C0C0}0.59} & \multicolumn{2}{c|}{\cellcolor[HTML]{C0C0C0}0.59} & \multicolumn{2}{c|}{\cellcolor[HTML]{C0C0C0}0.61} & \multicolumn{2}{c|}{-0.13} & \multicolumn{2}{c|}{0.29} & \multicolumn{2}{c|}{0.30} \\
\multicolumn{1}{|c|}{\cellcolor[HTML]{EFEFEF}} & \multicolumn{1}{l|}{cocomo81} & \multicolumn{2}{c|}{0.10} & \multicolumn{2}{c|}{0.05} & \multicolumn{2}{c|}{0.18} & \multicolumn{2}{c|}{0.15} & \multicolumn{2}{c|}{\cellcolor[HTML]{C0C0C0}0.27} & \multicolumn{2}{c|}{\cellcolor[HTML]{C0C0C0}0.25} & \multicolumn{2}{c|}{-1.14} & \multicolumn{2}{c|}{0.20} & \multicolumn{2}{c|}{\cellcolor[HTML]{C0C0C0}0.27} \\
\multicolumn{1}{|c|}{\multirow{-3}{*}{\cellcolor[HTML]{EFEFEF}\textbf{cocomo}}} & \multicolumn{1}{l|}{nasa93} & \multicolumn{2}{c|}{0.08} & \multicolumn{2}{c|}{0.14} & \multicolumn{2}{c|}{0.36} & \multicolumn{2}{c|}{0.37} & \multicolumn{2}{c|}{0.36} & \multicolumn{2}{c|}{\cellcolor[HTML]{C0C0C0}0.41} & \multicolumn{2}{c|}{0.34} & \multicolumn{2}{c|}{\cellcolor[HTML]{C0C0C0}0.41} & \multicolumn{2}{c|}{0.30} 
\end{tabular}

\end{adjustbox}
\end{table*}

\subsection{Performance Metrics}

The results from each test set are evaluated in terms  Magnitude of the Relative Error (MRE) and Standardized Accuracy (SA). 
MRE is defined in terms of 
AR,  the magnitude of the absolute residual. This is  computed from the difference between predicted and actual effort values:
\[
\mathit{AR} = |\mathit{actual}_i - \mathit{predicted}_i|
\] 
MRE is the magnitude of the relative error calculated by expressing AR as a ratio of   actual effort:
\[
\mathit{MRE} = \frac{|\mathit{actual}_i - \mathit{predicted}_i|}{\mathit{actual}_i}
\]

MRE is criticized by some researchers as it is biased towards error underestimations~\cite{foss2003simulation,kitchenham2001accuracy,korte2008confidence,port2008comparative,shepperd2000building,stensrud2003further}. Nevertheless, we use it here
since  there exists known baselines for human performance in effort estimation expressed in terms of MRE~\cite{Jorgensen03}. 
 
Because of the issues with MRE, some researchers prefer other (more standardized) measurements, such as  Standardized Accuracy (SA)~\cite{langdon2016exact,shepperd2012evaluating}.
SA is based on Mean Absolute Error (MAE), which is defined in terms of 
\[
\mathit{MAE}=\frac{1}{N}\sum_{i=1}^n|\mathit{RealEffort}_i-\mathit{EstimatedEffort}_i|
\]
where $N$ is the number of projects used for evaluating the performance. SA uses MAE as follows:
\[
\mathit{SA} = (1-\frac{\mathit{MAE}_{P_{j}}}{\mathit{MAE}_{r_{guess}}})\times 100
\]
where $\mathit{MAE}_{P_{j}}$ is the MAE of the approach $P_j$ being evaluated and $\mathit{MAE}_{r_{\mathit{guess}}}$ is the MAE of a large number (e.g., 1000 runs) of random guesses. 
Over many runs,  $\mathit{MAE}_{r_{\mathit{guess}}}$ will converge on simply using the sample mean~\cite{shepperd2012evaluating}. That is, SA represents how much better $P_j$ is than random guessing. Values near zero means that the prediction model $P_j$ is practically useless, performing little better than  random guesses~\cite{shepperd2012evaluating}.

Note that for MRE values, {\em smaller} are {\em better} and for SA values, {\em larger} are {\em better}.
We use these since there are advocates for both in the literature.
For example, Shepperd and MacDonell argue convincingly for the use of
 SA~\cite{shepperd2012evaluating} (as well as for the use of effect size tests in effort estimation).
 Also in 2016, MRE was used by Sarro et al.~\cite{sarro2016multi} to argue their estimators were competitive with human estimates
 (which Molokken et al.~\cite{molokken2003review} says lies within 30\% and 40\% of the true value).

\subsection{Statistical Methods}

From the cross-valuations,
we  report the {\em median} value, which is the 50th percentile of the   test scores seen in the {\em M*N results}. For each data set, the results from a {\em M*N-way} are sorted by their {\em  median} value, then {\em ranked} using the Scott-Knott test
recommended for ranking effort estimation experiments by Mittas et al. in TSE'13~\cite{Mittas13}. 

Scott-Knott is a top-down bi-clustering
method that recursively divides sorted treatments. Division stops when there is only one treatment left or when a division of numerous treatments generates splits that are statistically {\em indistinguishable}. 
To judge when two sets of treatments are indistinguishable, we use a conjunction of {\em both}  a 95\% bootstrap significance test~\cite{efron93} {\em and}
a A12 test for a non-small effect size difference in the distributions~\cite{MenziesNeg:2017}. These tests were used since their non-parametric nature avoids issues with non-Gaussian
distributions.

\begin{table*}[!t]
\centering
\caption{MRE (Magnitude of the Relative Error) of contemporary data sets (Part 1), \underline{lower} values are \underline{better}. Same format as Table~\ref{table:result_mre}; }

\label{table:result_mre_con_1}
\begin{adjustbox}{max width=0.87\textwidth}
\centering
\begin{tabular}{|c|l|cccccccc|}
\hline
\rowcolor[HTML]{EFEFEF} 
\multicolumn{2}{|c|}{\cellcolor[HTML]{EFEFEF}} & \multicolumn{4}{c|}{\cellcolor[HTML]{EFEFEF}\textbf{Scikit-Learn}} & \multicolumn{2}{c|}{\cellcolor[HTML]{EFEFEF}\textbf{Tuned}} & \multicolumn{2}{c|}{\cellcolor[HTML]{EFEFEF}\textbf{Other Methods}} \\ \cline{3-10} 
\multicolumn{2}{|c|}{\multirow{-2}{*}{\cellcolor[HTML]{EFEFEF}\textbf{Dataset}}} & \multicolumn{1}{c|}{\textbf{KNN}} & \multicolumn{1}{c|}{\textbf{SVR}} & \multicolumn{1}{c|}{\textbf{CART}} & \multicolumn{1}{c|}{\textbf{RF}} & \multicolumn{1}{c|}{\textbf{CART\_DE}} & \multicolumn{1}{c|}{\textbf{ROME}} & \multicolumn{1}{c|}{\textbf{ATLM}} & \multicolumn{1}{c|}{\textbf{LP4EE}} \\ \hline
\cellcolor[HTML]{EFEFEF} & abrash-black-book   & 0.98 & 0.81 & 0.33 & 0.96 & \cellcolor[HTML]{C0C0C0}0.08 & \cellcolor[HTML]{C0C0C0}0.08 & 0.61 & 0.62 \\
\cellcolor[HTML]{EFEFEF} & absinthe   & 9.99 & 11.69 & 13.99 & 33.99 & \cellcolor[HTML]{C0C0C0}6.86 & \cellcolor[HTML]{C0C0C0}5.82 & 16.58 & 8.68 \\
\cellcolor[HTML]{EFEFEF} & android-maps-utils   & 0.44 & 0.38 & 0.33 & 0.33 & \cellcolor[HTML]{C0C0C0}0.02 & \cellcolor[HTML]{C0C0C0}0.01 & 0.32 & 0.25 \\
\cellcolor[HTML]{EFEFEF} & AngleSharp   & \cellcolor[HTML]{C0C0C0}2.89 & 6.68 & 16.67 & 13.33 & 4.76 & 3.49 & 17.2 & 6.52 \\
\cellcolor[HTML]{EFEFEF} & aws-vault   & 0.36 & 0.67 & 0.86 & 0.41 & \cellcolor[HTML]{C0C0C0}0.01 & 0.14 & 0.54 & 0.18 \\
\cellcolor[HTML]{EFEFEF} & bisq   & 10.33 & 16.87 & \cellcolor[HTML]{C0C0C0}1.57 & 9.48 & 4.09 & 2.86 & 5.67 & 6.65 \\
\cellcolor[HTML]{EFEFEF} & bootstrap-tagsinput   & 2.03 & 0.61 & 3.69 & 2.19 & \cellcolor[HTML]{C0C0C0}0.01 & 0.58 & 2.51 & 0.8 \\
\cellcolor[HTML]{EFEFEF} & bosun   & 0.25 & 3.02 & \cellcolor[HTML]{C0C0C0}0.01 & 2.99 & 1.42 & 0.74 & 0.81 & 1.23 \\
\cellcolor[HTML]{EFEFEF} & chromedeveditor   & 17.67 & 32.57 & \cellcolor[HTML]{C0C0C0}0.01 & 3.99 & 0.22 & 0.22 & 4.67 & 6.04 \\
\cellcolor[HTML]{EFEFEF} & ckeditor5   & 3.47 & 4.88 & 0.59 & 1.79 & \cellcolor[HTML]{C0C0C0}0.33 & \cellcolor[HTML]{C0C0C0}0.33 & 1.59 & 1.97 \\
\cellcolor[HTML]{EFEFEF} & CotEditor   & 0.33 & 0.17 & 0.65 & 0.64 & \cellcolor[HTML]{C0C0C0}0.11 & \cellcolor[HTML]{C0C0C0}0.11 & 0.57 & 0.17 \\
\cellcolor[HTML]{EFEFEF} & cowrie   & 7.83 & 6.15 & 1.62 & \cellcolor[HTML]{C0C0C0}1.57 & \cellcolor[HTML]{C0C0C0}1.52 & \cellcolor[HTML]{C0C0C0}1.52 & 2.44 & 2.46 \\
\cellcolor[HTML]{EFEFEF} & cssicon   & \cellcolor[HTML]{C0C0C0}0.01 & 0.14 & \cellcolor[HTML]{C0C0C0}0.01 & 16.33 & 0.64 & 0.64 & 1.46 & 2.65 \\
\cellcolor[HTML]{EFEFEF} & ctf-wiki   & 5.44 & 7.48 & 3.33 & 4.44 & \cellcolor[HTML]{C0C0C0}1.63 & \cellcolor[HTML]{C0C0C0}1.73 & 4.94 & 3.29 \\
\cellcolor[HTML]{EFEFEF} & devd   & 0.71 & 0.65 & 0.11 & 0.22 & 0.07 & \cellcolor[HTML]{C0C0C0}0.01 & 0.21 & 0.28 \\
\cellcolor[HTML]{EFEFEF} & dynet   & 2.49 & 15.27 & 1.49 & 2.99 & \cellcolor[HTML]{C0C0C0}1.07 & \cellcolor[HTML]{C0C0C0}1.07 & 3.92 & 4.38 \\
\cellcolor[HTML]{EFEFEF} & electron-sample-apps   & 0.49 & \cellcolor[HTML]{C0C0C0}0.34 & 0.49 & 3.17 & \cellcolor[HTML]{C0C0C0}0.35 & 1.44 & 0.6 & 0.49 \\
\cellcolor[HTML]{EFEFEF} & embark   & 0.24 & 0.21 & \cellcolor[HTML]{C0C0C0}0.11 & 0.16 & \cellcolor[HTML]{C0C0C0}0.09 & \cellcolor[HTML]{C0C0C0}0.09 & 0.17 & 0.13 \\
\cellcolor[HTML]{EFEFEF} & ethminer   & \cellcolor[HTML]{C0C0C0}0.33 & 42.81 & 1.99 & 6.83 & 3.29 & 2.75 & 10.9 & 14.14 \\
\cellcolor[HTML]{EFEFEF} & f2etest   & 2.99 & 1.71 & 9.99 & 4.33 & \cellcolor[HTML]{C0C0C0}0.64 & \cellcolor[HTML]{C0C0C0}0.56 & 8.52 & 1.69 \\
\cellcolor[HTML]{EFEFEF} & fis   & 7.17 & 0.25 & 0.99 & 0.67 & \cellcolor[HTML]{C0C0C0}0.06 & 0.29 & 0.78 & 0.35 \\
\cellcolor[HTML]{EFEFEF} & flask\_jsondash   & \cellcolor[HTML]{C0C0C0}1.33 & 3.32 & 29.99 & 47.99 & 18.77 & 21.57 & 29.56 & 21.1 \\
\cellcolor[HTML]{EFEFEF} & frozenui   & 0.99 & 1.16 & 9.99 & 3.17 & 0.65 & \cellcolor[HTML]{C0C0C0}0.19 & 8.02 & 0.31 \\
\cellcolor[HTML]{EFEFEF} & goby   & 0.57 & \cellcolor[HTML]{C0C0C0}0.02 & 2.05 & 1.51 & \cellcolor[HTML]{C0C0C0}0.05 & 0.21 & 1.73 & 0.44 \\
\cellcolor[HTML]{EFEFEF} & gosec   & 1.17 & 4.02 & \cellcolor[HTML]{C0C0C0}0.01 & 1.33 & 0.77 & 0.67 & 0.72 & 0.99 \\
\cellcolor[HTML]{EFEFEF} & gradle-play-publisher   & 2.13 & 0.57 & 0.39 & 1.87 & \cellcolor[HTML]{C0C0C0}0.04 & \cellcolor[HTML]{C0C0C0}0.11 & 0.43 & 0.25 \\
\cellcolor[HTML]{EFEFEF} & HackingWithSwift   & 0.52 & 0.61 & 0.43 & 0.38 & \cellcolor[HTML]{C0C0C0}0.33 & \cellcolor[HTML]{C0C0C0}0.33 & 0.49 & 0.37 \\
\cellcolor[HTML]{EFEFEF} & hadolint   & 1.17 & 3.39 & 8.49 & 6.67 & 0.74 & \cellcolor[HTML]{C0C0C0}0.12 & 7.8 & 1.37 \\
\cellcolor[HTML]{EFEFEF} & horizon   & 2.99 & 9.49 & \cellcolor[HTML]{C0C0C0}0.99 & 1.99 & 1.33 & 1.33 & 2.13 & 2.87 \\
\cellcolor[HTML]{EFEFEF} & ImageOptim-CLI   & 0.33 & 0.29 & \cellcolor[HTML]{C0C0C0}0.01 & 0.22 & 0.22 & \cellcolor[HTML]{C0C0C0}0.01 & \cellcolor[HTML]{C0C0C0}0.04 & \cellcolor[HTML]{C0C0C0}0.05 \\
\cellcolor[HTML]{EFEFEF} & javacpp   & 4.67 & 4.19 & 5.99 & 6.22 & \cellcolor[HTML]{C0C0C0}2.48 & \cellcolor[HTML]{C0C0C0}2.42 & 4.69 & 3.41 \\
\cellcolor[HTML]{EFEFEF} & KSCrash   & 2.33 & 2.18 & 0.99 & 1.17 & \cellcolor[HTML]{C0C0C0}0.02 & \cellcolor[HTML]{C0C0C0}0.02 & 1.03 & 0.62 \\
\cellcolor[HTML]{EFEFEF} & LayoutKit   & 0.67 & 0.26 & 3.33 & 4.22 & \cellcolor[HTML]{C0C0C0}0.04 & \cellcolor[HTML]{C0C0C0}0.03 & 2.63 & 0.76 \\
\cellcolor[HTML]{EFEFEF} & LeafPic   & 4.33 & 1.29 & \cellcolor[HTML]{C0C0C0}0.01 & 0.99 & \cellcolor[HTML]{C0C0C0}0.01 & \cellcolor[HTML]{C0C0C0}0.01 & 0.18 & 0.4 \\
\cellcolor[HTML]{EFEFEF} & lossless-cut   & 0.87 & 0.71 & 0.67 & \cellcolor[HTML]{C0C0C0}0.16 & \cellcolor[HTML]{C0C0C0}0.06 & \cellcolor[HTML]{C0C0C0}0.06 & 0.56 & \cellcolor[HTML]{C0C0C0}0.1 \\
\cellcolor[HTML]{EFEFEF} & material   & 22.22 & \cellcolor[HTML]{C0C0C0}6.07 & 16.67 & 20.56 & 12.57 & 12.57 & 20.84 & 13.7 \\
\cellcolor[HTML]{EFEFEF} & mkdocs-material   & 6.67 & 22.57 & \cellcolor[HTML]{C0C0C0}0.01 & 1.99 & 0.33 & 0.33 & 2.7 & 4.44 \\
\cellcolor[HTML]{EFEFEF} & mobile-angular-ui   & 1.33 & 0.52 & \cellcolor[HTML]{C0C0C0}0.01 & 2.11 & \cellcolor[HTML]{C0C0C0}0.02 & \cellcolor[HTML]{C0C0C0}0.01 & 0.78 & 1.23 \\
\cellcolor[HTML]{EFEFEF} & moco   & 0.12 & 0.09 & 0.15 & 0.12 & 0.03 & 0.04 & 0.09 & \cellcolor[HTML]{C0C0C0}0.01 \\
\cellcolor[HTML]{EFEFEF} & neo   & 0.12 & 0.61 & 0.11 & 0.17 & \cellcolor[HTML]{C0C0C0}0.02 & \cellcolor[HTML]{C0C0C0}0.02 & 0.21 & 0.22 \\
\cellcolor[HTML]{EFEFEF} & open-location-code   & 4.33 & 3.65 & \cellcolor[HTML]{C0C0C0}0.99 & 2.33 & \cellcolor[HTML]{C0C0C0}0.99 & \cellcolor[HTML]{C0C0C0}0.99 & 1.52 & 1.94 \\
\cellcolor[HTML]{EFEFEF} & osv   & \cellcolor[HTML]{C0C0C0}10.99 & 18.99 & 501.99 & 322.33 & \cellcolor[HTML]{C0C0C0}11.32 & 57.32 & 456.08 & 87.85 \\
\cellcolor[HTML]{EFEFEF} & patroni   & 0.07 & 0.12 & 0.11 & 0.06 & 0.06 & \cellcolor[HTML]{C0C0C0}0.01 & 0.12 & 0.14 \\
\cellcolor[HTML]{EFEFEF} & pigeon-maps   & 0.33 & 0.38 & 0.99 & 0.99 & \cellcolor[HTML]{C0C0C0}0.09 & \cellcolor[HTML]{C0C0C0}0.17 & 0.99 & 0.48 \\
\cellcolor[HTML]{EFEFEF} & polaris-react   & \cellcolor[HTML]{C0C0C0}0.99 & 121.21 & 4.99 & 4.99 & 5.99 & 5.29 & 26.48 & 34.46 \\
\cellcolor[HTML]{EFEFEF} & polr   & 15.33 & 11.58 & 43.99 & 29.67 & \cellcolor[HTML]{C0C0C0}4.54 & 16.99 & 45.61 & 19.77 \\
\cellcolor[HTML]{EFEFEF} & react-flip-move   & 1.33 & 0.89 & 5.99 & 1.99 & \cellcolor[HTML]{C0C0C0}0.35 & 0.88 & 4.14 & 0.74 \\
\cellcolor[HTML]{EFEFEF} & react-native-gesture-handler   & 0.11 & 0.73 & \cellcolor[HTML]{C0C0C0}0.01 & 0.99 & 0.36 & 0.38 & 0.15 & 0.23 \\
\cellcolor[HTML]{EFEFEF} & react-overdrive   & 2.99 & 0.91 & 0.49 & 0.49 & \cellcolor[HTML]{C0C0C0}0.24 & \cellcolor[HTML]{C0C0C0}0.29 & 0.69 & 0.44 \\
\cellcolor[HTML]{EFEFEF} & ring   & 0.33 & 0.64 & 0.79 & 0.67 & 0.65 & \cellcolor[HTML]{C0C0C0}0.19 & 0.6 & \cellcolor[HTML]{C0C0C0}0.23 \\
\cellcolor[HTML]{EFEFEF} & SCRecorder   & 0.83 & 2.23 & 1.99 & 2.33 & 0.25 & \cellcolor[HTML]{C0C0C0}0.08 & 1.62 & 0.84 \\
\cellcolor[HTML]{EFEFEF} & Solve-App-Store-Review-Problem   & 3.33 & 0.92 & 7.99 & 5.67 & \cellcolor[HTML]{C0C0C0}0.09 & 0.24 & 7.58 & 1.33 \\
\cellcolor[HTML]{EFEFEF} & sqldelight   & 0.28 & 0.45 & \cellcolor[HTML]{C0C0C0}0.03 & 0.26 & \cellcolor[HTML]{C0C0C0}0.02 & \cellcolor[HTML]{C0C0C0}0.02 & 0.08 & 0.08 \\
\cellcolor[HTML]{EFEFEF} & tabulator   & 2.44 & 2.99 & 0.78 & 0.99 & \cellcolor[HTML]{C0C0C0}0.01 & \cellcolor[HTML]{C0C0C0}0.01 & 1.08 & 1.08 \\
\cellcolor[HTML]{EFEFEF} & TheAmazingAudioEngine   & \cellcolor[HTML]{C0C0C0}0.02 & \cellcolor[HTML]{C0C0C0}0.02 & 3.49 & 0.17 & 0.25 & \cellcolor[HTML]{C0C0C0}0.07 & 2.24 & 0.38 \\
\cellcolor[HTML]{EFEFEF} & ts-jest   & 0.27 & 0.32 & 0.52 & 0.27 & \cellcolor[HTML]{C0C0C0}0.05 & 0.19 & 0.42 & 0.11 \\
\cellcolor[HTML]{EFEFEF} & vscode-cpptools   & \cellcolor[HTML]{C0C0C0}0.14 & 0.48 & 0.71 & 0.49 & 0.46 & 0.59 & 0.86 & 0.43 \\
\cellcolor[HTML]{EFEFEF} & vue-meta   & 1.17 & 4.64 & 1.99 & 1.83 & \cellcolor[HTML]{C0C0C0}0.49 & \cellcolor[HTML]{C0C0C0}0.49 & 2.44 & 2.29 \\
\cellcolor[HTML]{EFEFEF} & XcodeGen   & 3.14 & 7.04 & 0.99 & 1.95 & 2.33 & \cellcolor[HTML]{C0C0C0}0.67 & 2.37 & 2.51 \\
\multirow{-60}{*}{\cellcolor[HTML]{EFEFEF}\textbf{contemporary}} & XposedBridge   & \cellcolor[HTML]{C0C0C0}0.01 & 0.86 & 0.75 & 0.11 & \cellcolor[HTML]{C0C0C0}0.04 & \cellcolor[HTML]{C0C0C0}0.04 & 0.63 & 0.22 \\ \hline
\multicolumn{2}{|c|}{\multirow{-1}{*}{\cellcolor[HTML]{EFEFEF}\textbf{Win Time Count}}} & \multicolumn{1}{c|}{\textbf{9}} & \multicolumn{1}{c|}{\textbf{4}} & \multicolumn{1}{c|}{\textbf{14}} & \multicolumn{1}{c|}{\textbf{2}} & \multicolumn{1}{c|}{\textbf{36}} & \multicolumn{1}{c|}{\textbf{35}} & \multicolumn{1}{c|}{\textbf{1}} & \multicolumn{1}{c|}{\textbf{4}} \\ \hline
\end{tabular}

\end{adjustbox}
\end{table*}

\begin{table*}[!t]
\centering
\caption{MRE (Magnitude of the Relative Error) of contemporary data sets (Part 2), \underline{lower} values are \underline{better}. Same format as Table~\ref{table:result_mre}; }
\label{table:result_mre_con_2}
\begin{adjustbox}{max width=0.87\textwidth}
\centering
\begin{tabular}{|c|l|cccccccc|}
\hline
\rowcolor[HTML]{EFEFEF} 
\multicolumn{2}{|c|}{\cellcolor[HTML]{EFEFEF}} & \multicolumn{4}{c|}{\cellcolor[HTML]{EFEFEF}\textbf{Scikit-Learn}} & \multicolumn{2}{c|}{\cellcolor[HTML]{EFEFEF}\textbf{Tuned}} & \multicolumn{2}{c|}{\cellcolor[HTML]{EFEFEF}\textbf{Other Methods}} \\ \cline{3-10} 
\multicolumn{2}{|c|}{\multirow{-2}{*}{\cellcolor[HTML]{EFEFEF}\textbf{Dataset}}} & \multicolumn{1}{c|}{\textbf{KNN}} & \multicolumn{1}{c|}{\textbf{SVR}} & \multicolumn{1}{c|}{\textbf{CART}} & \multicolumn{1}{c|}{\textbf{RF}} & \multicolumn{1}{c|}{\textbf{CART\_DE}} & \multicolumn{1}{c|}{\textbf{ROME}} & \multicolumn{1}{c|}{\textbf{ATLM}} & \multicolumn{1}{c|}{\textbf{LP4EE}} \\ \hline
\cellcolor[HTML]{EFEFEF} & aeron   & 25.17 & \cellcolor[HTML]{C0C0C0}16.95 & 30.49 & 21.49 & \cellcolor[HTML]{C0C0C0}18.43 & \cellcolor[HTML]{C0C0C0}14.07 & 29.27 & \cellcolor[HTML]{C0C0C0}16.85 \\
\cellcolor[HTML]{EFEFEF} & alasql   & 1.22 & 0.23 & 0.67 & 0.48 & 0.12 & \cellcolor[HTML]{C0C0C0}0.01 & 0.86 & 0.82 \\
\cellcolor[HTML]{EFEFEF} & android-advancedrecyclerview   & 0.74 & 0.13 & \cellcolor[HTML]{C0C0C0}0.01 & 0.07 & 0.27 & 0.12 & 0.11 & 0.11 \\
\cellcolor[HTML]{EFEFEF} & Android-Image-Cropper   & 8.22 & \cellcolor[HTML]{C0C0C0}1.08 & 15.67 & 11.44 & \cellcolor[HTML]{C0C0C0}1.11 & \cellcolor[HTML]{C0C0C0}1.11 & 9.97 & \cellcolor[HTML]{C0C0C0}1.77 \\
\cellcolor[HTML]{EFEFEF} & AndroidPicker   & 2.33 & \cellcolor[HTML]{C0C0C0}0.98 & 2.99 & 1.99 & 2.43 & 2.69 & 2.71 & 1.99 \\
\cellcolor[HTML]{EFEFEF} & angular-seed   & 2.67 & 7.69 & 0.42 & 1.67 & 1.43 & \cellcolor[HTML]{C0C0C0}0.33 & 1.68 & 1.95 \\
\cellcolor[HTML]{EFEFEF} & BackstopJS   & \cellcolor[HTML]{C0C0C0}0.67 & 0.69 & 1.33 & 1.48 & 0.84 & 0.84 & 1.84 & 0.93 \\
\cellcolor[HTML]{EFEFEF} & brave-browser   & 0.42 & 1.44 & 3.55 & 0.52 & \cellcolor[HTML]{C0C0C0}0.03 & \cellcolor[HTML]{C0C0C0}0.03 & 2.27 & 0.34 \\
\cellcolor[HTML]{EFEFEF} & cachecloud   & 1.78 & \cellcolor[HTML]{C0C0C0}0.01 & 9.67 & 4.99 & 1.38 & 1.27 & 8.17 & 1.81 \\
\cellcolor[HTML]{EFEFEF} & caprine   & 0.33 & 0.25 & 0.37 & 0.33 & \cellcolor[HTML]{C0C0C0}0.01 & \cellcolor[HTML]{C0C0C0}0.02 & 0.25 & \cellcolor[HTML]{C0C0C0}0.05 \\
\cellcolor[HTML]{EFEFEF} & cfssl   & 0.33 & 0.73 & 0.49 & \cellcolor[HTML]{C0C0C0}0.17 & \cellcolor[HTML]{C0C0C0}0.06 & \cellcolor[HTML]{C0C0C0}0.12 & 0.33 & \cellcolor[HTML]{C0C0C0}0.09 \\
\cellcolor[HTML]{EFEFEF} & clappr   & 1.49 & 9.31 & 0.25 & 0.49 & 0.85 & \cellcolor[HTML]{C0C0C0}0.06 & 3.16 & 4.09 \\
\cellcolor[HTML]{EFEFEF} & community-edition   & 0.22 & 3.35 & 0.33 & 0.44 & \cellcolor[HTML]{C0C0C0}0.04 & \cellcolor[HTML]{C0C0C0}0.01 & 0.7 & 0.57 \\
\cellcolor[HTML]{EFEFEF} & core   & 0.76 & 1.85 & 2.73 & 2.06 & \cellcolor[HTML]{C0C0C0}0.55 & \cellcolor[HTML]{C0C0C0}0.55 & 1.86 & 0.76 \\
\cellcolor[HTML]{EFEFEF} & cpprestsdk   & 4.08 & 1.62 & 0.75 & 1.58 & 0.37 & \cellcolor[HTML]{C0C0C0}0.03 & 0.74 & 0.34 \\
\cellcolor[HTML]{EFEFEF} & Dexie.js   & 7.99 & 7.15 & 4.49 & 7.67 & \cellcolor[HTML]{C0C0C0}2.56 & 3.79 & 6.25 & 4.57 \\
\cellcolor[HTML]{EFEFEF} & diesel   & 0.33 & 0.25 & 3.04 & 1.19 & \cellcolor[HTML]{C0C0C0}0.02 & \cellcolor[HTML]{C0C0C0}0.02 & 2.52 & 0.12 \\
\cellcolor[HTML]{EFEFEF} & discord.js   & 0.35 & 1.44 & 0.24 & 0.35 & \cellcolor[HTML]{C0C0C0}0.02 & \cellcolor[HTML]{C0C0C0}0.02 & 0.4 & 0.43 \\
\cellcolor[HTML]{EFEFEF} & documentation   & 0.25 & 1.48 & 0.25 & 0.42 & \cellcolor[HTML]{C0C0C0}0.11 & \cellcolor[HTML]{C0C0C0}0.09 & 0.5 & 0.44 \\
\cellcolor[HTML]{EFEFEF} & EarlGrey   & 0.99 & 0.68 & 0.99 & 0.67 & \cellcolor[HTML]{C0C0C0}0.21 & \cellcolor[HTML]{C0C0C0}0.19 & 0.95 & 0.5 \\
\cellcolor[HTML]{EFEFEF} & EIPs   & \cellcolor[HTML]{C0C0C0}0.08 & 0.21 & 0.44 & 0.54 & 0.25 & 0.25 & 0.39 & 0.22 \\
\cellcolor[HTML]{EFEFEF} & error-prone   & 15.33 & 26.22 & 0.99 & 10.33 & 2.79 & \cellcolor[HTML]{C0C0C0}0.49 & 5.6 & 7.41 \\
\cellcolor[HTML]{EFEFEF} & evil-icons   & 0.47 & 0.37 & 0.19 & \cellcolor[HTML]{C0C0C0}0.13 & \cellcolor[HTML]{C0C0C0}0.09 & \cellcolor[HTML]{C0C0C0}0.09 & 0.26 & 0.27 \\
\cellcolor[HTML]{EFEFEF} & exceljs   & 1.75 & 1.41 & 0.25 & 0.33 & \cellcolor[HTML]{C0C0C0}0.05 & \cellcolor[HTML]{C0C0C0}0.01 & 0.42 & 0.28 \\
\cellcolor[HTML]{EFEFEF} & fission   & 0.52 & 0.12 & 0.16 & 0.05 & \cellcolor[HTML]{C0C0C0}0.01 & \cellcolor[HTML]{C0C0C0}0.01 & 0.16 & 0.07 \\
\cellcolor[HTML]{EFEFEF} & flannel   & 0.12 & 0.04 & 0.21 & 0.07 & 0.05 & \cellcolor[HTML]{C0C0C0}0.01 & 0.17 & 0.08 \\
\cellcolor[HTML]{EFEFEF} & go-git   & 3.56 & 6.93 & 0.99 & 1.44 & \cellcolor[HTML]{C0C0C0}0.83 & \cellcolor[HTML]{C0C0C0}0.85 & 2.06 & 2.51 \\
\cellcolor[HTML]{EFEFEF} & goreleaser   & 1.26 & 2.65 & 0.38 & 1.21 & \cellcolor[HTML]{C0C0C0}0.29 & \cellcolor[HTML]{C0C0C0}0.29 & 0.89 & 0.97 \\
\cellcolor[HTML]{EFEFEF} & haven   & 1.18 & 0.23 & 0.09 & 0.36 & 0.07 & \cellcolor[HTML]{C0C0C0}0.01 & 0.24 & 0.24 \\
\cellcolor[HTML]{EFEFEF} & hospitalrun-frontend   & 2.08 & 2.98 & 1.08 & 3.28 & 1.11 & \cellcolor[HTML]{C0C0C0}0.65 & 1.57 & 1.56 \\
\cellcolor[HTML]{EFEFEF} & HTextView   & 1.33 & \cellcolor[HTML]{C0C0C0}0.07 & 0.99 & 1.99 & \cellcolor[HTML]{C0C0C0}0.08 & \cellcolor[HTML]{C0C0C0}0.12 & 0.82 & 0.47 \\
\cellcolor[HTML]{EFEFEF} & humhub   & \cellcolor[HTML]{C0C0C0}20.67 & 60.01 & 49.99 & 40.99 & 30.88 & 30.88 & 49.75 & 33.92 \\
\cellcolor[HTML]{EFEFEF} & huxpro.github.io   & \cellcolor[HTML]{C0C0C0}0.01 & 4.41 & 1.99 & 7.99 & 2.91 & 3.81 & 6.89 & 6.78 \\
\cellcolor[HTML]{EFEFEF} & incubator-tvm   & 2.09 & 9.83 & \cellcolor[HTML]{C0C0C0}1.43 & 2.09 & 2.36 & 1.86 & 3.22 & 4.03 \\
\cellcolor[HTML]{EFEFEF} & is-thirteen   & 0.49 & 0.55 & 0.59 & 0.43 & 0.44 & 0.43 & 0.65 & \cellcolor[HTML]{C0C0C0}0.36 \\
\cellcolor[HTML]{EFEFEF} & jellyfin   & \cellcolor[HTML]{C0C0C0}13.46 & 16.59 & 17.97 & 19.11 & 17.66 & 17.59 & 22.99 & 20.05 \\
\cellcolor[HTML]{EFEFEF} & kanboard   & 0.56 & 4.87 & \cellcolor[HTML]{C0C0C0}0.17 & 1.11 & \cellcolor[HTML]{C0C0C0}0.17 & 0.43 & 1.02 & 1.62 \\
\cellcolor[HTML]{EFEFEF} & kingshard   & 0.39 & 0.11 & 5.33 & 0.17 & \cellcolor[HTML]{C0C0C0}0.06 & 0.33 & 4.24 & 0.27 \\
\cellcolor[HTML]{EFEFEF} & lucida   & 0.33 & 2.55 & \cellcolor[HTML]{C0C0C0}0.01 & 1.99 & 0.39 & \cellcolor[HTML]{C0C0C0}0.01 & 0.56 & 0.7 \\
\cellcolor[HTML]{EFEFEF} & moon   & 0.22 & 0.86 & 0.67 & 0.17 & \cellcolor[HTML]{C0C0C0}0.06 & \cellcolor[HTML]{C0C0C0}0.03 & 0.46 & 0.18 \\
\cellcolor[HTML]{EFEFEF} & mybatis-generator-gui   & 2.33 & 3.62 & 0.74 & 0.67 & \cellcolor[HTML]{C0C0C0}0.26 & 0.57 & 1.37 & 1.35 \\
\cellcolor[HTML]{EFEFEF} & ngx-bootstrap   & 3.22 & 9.32 & 0.67 & 0.33 & \cellcolor[HTML]{C0C0C0}0.06 & \cellcolor[HTML]{C0C0C0}0.06 & 1.48 & 1.82 \\
\cellcolor[HTML]{EFEFEF} & or-tools   & \cellcolor[HTML]{C0C0C0}11.17 & 20.14 & \cellcolor[HTML]{C0C0C0}10.99 & 13.82 & 18.92 & \cellcolor[HTML]{C0C0C0}10.81 & 13.58 & 14.22 \\
\cellcolor[HTML]{EFEFEF} & oss-fuzz   & 8.73 & 11.91 & 10.19 & 9.19 & \cellcolor[HTML]{C0C0C0}8.45 & \cellcolor[HTML]{C0C0C0}8.45 & 11.71 & 8.95 \\
\cellcolor[HTML]{EFEFEF} & places   & 0.15 & 0.15 & 0.11 & 0.06 & \cellcolor[HTML]{C0C0C0}0.02 & 0.05 & 0.11 & 0.04 \\
\cellcolor[HTML]{EFEFEF} & pulsar   & 2.71 & 9.56 & \cellcolor[HTML]{C0C0C0}0.99 & 2.19 & \cellcolor[HTML]{C0C0C0}0.99 & \cellcolor[HTML]{C0C0C0}0.99 & 3.36 & 3.62 \\
\cellcolor[HTML]{EFEFEF} & react-autosuggest   & 0.99 & 5.02 & \cellcolor[HTML]{C0C0C0}0.01 & \cellcolor[HTML]{C0C0C0}0.01 & 0.09 & 0.34 & 1.07 & 1.44 \\
\cellcolor[HTML]{EFEFEF} & react-map-gl   & 0.86 & 0.46 & 0.43 & 0.33 & \cellcolor[HTML]{C0C0C0}0.12 & \cellcolor[HTML]{C0C0C0}0.12 & 0.59 & 0.34 \\
\cellcolor[HTML]{EFEFEF} & react-native-paper   & 0.23 & \cellcolor[HTML]{C0C0C0}0.09 & 1.09 & 0.29 & 0.18 & 0.18 & 0.75 & 0.26 \\
\cellcolor[HTML]{EFEFEF} & redex   & 15.99 & 19.89 & \cellcolor[HTML]{C0C0C0}7.67 & 10.56 & 9.83 & 9.83 & 16.06 & 13.79 \\
\cellcolor[HTML]{EFEFEF} & resilience4j   & 4.56 & 0.52 & 0.67 & 0.44 & 0.28 & \cellcolor[HTML]{C0C0C0}0.21 & 0.78 & 0.41 \\
\cellcolor[HTML]{EFEFEF} & SCLAlertView-Swift   & 0.33 & 1.95 & 0.99 & 0.67 & \cellcolor[HTML]{C0C0C0}0.16 & 0.29 & 1.01 & 0.44 \\
\cellcolor[HTML]{EFEFEF} & sdk   & 13.83 & 15.29 & 13.26 & 13.29 & 13.26 & \cellcolor[HTML]{C0C0C0}11.26 & 13.58 & 12.33 \\
\cellcolor[HTML]{EFEFEF} & single-spa   & 0.89 & 1.39 & 0.33 & 0.56 & 0.17 & \cellcolor[HTML]{C0C0C0}0.01 & 0.44 & 0.35 \\
\cellcolor[HTML]{EFEFEF} & Squirrel.Windows   & 0.33 & 0.23 & 0.67 & 0.44 & \cellcolor[HTML]{C0C0C0}0.09 & 0.18 & 0.55 & 0.26 \\
\cellcolor[HTML]{EFEFEF} & thingsboard   & 0.43 & 0.22 & 1.69 & 0.47 & \cellcolor[HTML]{C0C0C0}0.09 & \cellcolor[HTML]{C0C0C0}0.13 & 1.42 & \cellcolor[HTML]{C0C0C0}0.16 \\
\cellcolor[HTML]{EFEFEF} & TranslationPlugin   & 1.22 & \cellcolor[HTML]{C0C0C0}0.23 & 0.76 & 1.33 & \cellcolor[HTML]{C0C0C0}0.16 & 1.43 & 1.03 & 1.29 \\
\cellcolor[HTML]{EFEFEF} & translations   & 0.56 & 0.69 & 9.99 & 2.67 & \cellcolor[HTML]{C0C0C0}0.37 & 0.94 & 6.95 & 1.44 \\
\cellcolor[HTML]{EFEFEF} & vue-multiselect   & 8.33 & 8.45 & 6.99 & 13.99 & \cellcolor[HTML]{C0C0C0}5.58 & 6.46 & 12.32 & 9.08 \\
\multirow{-60}{*}{\cellcolor[HTML]{EFEFEF}\textbf{contemporary}} & z3   & 11.73 & 22.25 & 6.49 & 6.65 & 9.67 & \cellcolor[HTML]{C0C0C0}5.92 & 12.77 & 9.35 \\ \hline
\multicolumn{2}{|c|}{\multirow{-1}{*}{\cellcolor[HTML]{EFEFEF}\textbf{Win Time Count}}} & \multicolumn{1}{c|}{\textbf{6}} & \multicolumn{1}{c|}{\textbf{7}} & \multicolumn{1}{c|}{\textbf{8}} & \multicolumn{1}{c|}{\textbf{3}} & \multicolumn{1}{c|}{\textbf{33}} & \multicolumn{1}{c|}{\textbf{37}} & \multicolumn{1}{c|}{\textbf{0}} & \multicolumn{1}{c|}{\textbf{6}} \\ \hline
\end{tabular}

\end{adjustbox}
\end{table*}

\begin{table*}[!t]
\centering
\caption{SA (Standard Accuracy) of contemporary data sets (Part 1), \underline{higher} values are \underline{better}. Same format as Table~\ref{table:result_sa}; }
\label{table:result_sa_con_1}
\begin{adjustbox}{max width=0.9\textwidth}
\centering
\begin{tabular}{|c|l|cccccccc|}
\hline
\rowcolor[HTML]{EFEFEF} 
\multicolumn{2}{|c|}{\cellcolor[HTML]{EFEFEF}} & \multicolumn{4}{c|}{\cellcolor[HTML]{EFEFEF}\textbf{Scikit-Learn}} & \multicolumn{2}{c|}{\cellcolor[HTML]{EFEFEF}\textbf{Tuned}} & \multicolumn{2}{c|}{\cellcolor[HTML]{EFEFEF}\textbf{Other Methods}} \\ \cline{3-10} 
\multicolumn{2}{|c|}{\multirow{-2}{*}{\cellcolor[HTML]{EFEFEF}\textbf{Dataset}}} & \multicolumn{1}{c|}{\textbf{KNN}} & \multicolumn{1}{c|}{\textbf{SVR}} & \multicolumn{1}{c|}{\textbf{CART}} & \multicolumn{1}{c|}{\textbf{RF}} & \multicolumn{1}{c|}{\textbf{CART\_DE}} & \multicolumn{1}{c|}{\textbf{ROME}} & \multicolumn{1}{c|}{\textbf{ATLM}} & \multicolumn{1}{c|}{\textbf{LP4EE}} \\ \hline
\cellcolor[HTML]{EFEFEF} & abrash-black-book    & 0.01 & 0.21 & 0.67 & -1.52 & \cellcolor[HTML]{C0C0C0}0.92 & \cellcolor[HTML]{C0C0C0}0.92 & 0.37 & 0.36 \\
\cellcolor[HTML]{EFEFEF} & absinthe    & -8.99 & -10.69 & -12.99 & -12.99 & -5.86 & \cellcolor[HTML]{C0C0C0}-4.64 & -15.58 & -7.66 \\
\cellcolor[HTML]{EFEFEF} & android-maps-utils    & 0.56 & 0.62 & 0.67 & 0.78 & 0.85 & \cellcolor[HTML]{C0C0C0}0.99 & 0.66 & 0.78 \\
\cellcolor[HTML]{EFEFEF} & AngleSharp    & \cellcolor[HTML]{C0C0C0}-1.89 & -5.68 & -12.33 & -12.33 & -5.59 & -3.76 & -16.2 & -5.49 \\
\cellcolor[HTML]{EFEFEF} & aws-vault    & 0.64 & 0.32 & 0.04 & 0.65 & \cellcolor[HTML]{C0C0C0}0.94 & 0.86 & 0.44 & 0.8 \\
\cellcolor[HTML]{EFEFEF} & bisq    & -9.33 & -15.87 & -2.99 & -10.52 & \cellcolor[HTML]{C0C0C0}-1.86 & \cellcolor[HTML]{C0C0C0}-1.86 & -4.69 & -5.65 \\
\cellcolor[HTML]{EFEFEF} & bootstrap-tagsinput    & -1.03 & 0.39 & -2.69 & -0.31 & \cellcolor[HTML]{C0C0C0}0.98 & 0.29 & -1.53 & 0.19 \\
\cellcolor[HTML]{EFEFEF} & bosun    & 0.75 & -2.02 & \cellcolor[HTML]{C0C0C0}0.99 & 0.08 & -0.42 & 0.25 & 0.22 & -0.21 \\
\cellcolor[HTML]{EFEFEF} & chromedeveditor    & -16.67 & -31.57 & \cellcolor[HTML]{C0C0C0}0.99 & -2.33 & 0.78 & 0.78 & -3.7 & -5.01 \\
\cellcolor[HTML]{EFEFEF} & ckeditor5    & -2.47 & -3.88 & 0.41 & -2.53 & 0.35 & \cellcolor[HTML]{C0C0C0}0.92 & -0.58 & -0.96 \\
\cellcolor[HTML]{EFEFEF} & CotEditor    & 0.67 & \cellcolor[HTML]{C0C0C0}0.83 & 0.08 & 0.66 & \cellcolor[HTML]{C0C0C0}0.89 & \cellcolor[HTML]{C0C0C0}0.89 & 0.44 & \cellcolor[HTML]{C0C0C0}0.85 \\
\cellcolor[HTML]{EFEFEF} & cowrie    & -6.83 & -5.15 & -0.63 & -1.98 & \cellcolor[HTML]{C0C0C0}-0.52 & -0.63 & -1.47 & -1.47 \\
\cellcolor[HTML]{EFEFEF} & cssicon    & \cellcolor[HTML]{C0C0C0}0.99 & 0.86 & \cellcolor[HTML]{C0C0C0}0.99 & -2.67 & -1.71 & -2.43 & -0.46 & -1.62 \\
\cellcolor[HTML]{EFEFEF} & ctf-wiki    & -4.44 & -6.48 & -6.99 & -5.22 & -0.73 & \cellcolor[HTML]{C0C0C0}-0.63 & -3.97 & -2.3 \\
\cellcolor[HTML]{EFEFEF} & devd    & 0.29 & 0.35 & 0.89 & 0.74 & 0.93 & \cellcolor[HTML]{C0C0C0}0.99 & 0.77 & 0.73 \\
\cellcolor[HTML]{EFEFEF} & dynet    & -1.49 & -14.27 & \cellcolor[HTML]{C0C0C0}0.99 & -1.51 & -0.51 & -0.07 & -2.93 & -3.41 \\
\cellcolor[HTML]{EFEFEF} & electron-sample-apps    & 0.51 & 0.66 & 0.51 & -0.83 & 0.65 & \cellcolor[HTML]{C0C0C0}0.75 & 0.39 & 0.54 \\
\cellcolor[HTML]{EFEFEF} & embark    & 0.76 & 0.79 & 0.68 & 0.83 & \cellcolor[HTML]{C0C0C0}0.91 & \cellcolor[HTML]{C0C0C0}0.91 & 0.81 & 0.85 \\
\cellcolor[HTML]{EFEFEF} & ethminer    & \cellcolor[HTML]{C0C0C0}0.67 & -41.81 & -2.01 & -9.33 & -1.01 & -1.75 & -9.88 & -13.11 \\
\cellcolor[HTML]{EFEFEF} & f2etest    & -1.99 & -0.71 & -8.99 & -4.99 & \cellcolor[HTML]{C0C0C0}0.26 & \cellcolor[HTML]{C0C0C0}0.26 & -7.55 & -0.7 \\
\cellcolor[HTML]{EFEFEF} & fis    & -6.17 & 0.75 & 0.01 & 0.33 & \cellcolor[HTML]{C0C0C0}0.94 & 0.71 & 0.23 & 0.66 \\
\cellcolor[HTML]{EFEFEF} & flask\_jsondash    & \cellcolor[HTML]{C0C0C0}-0.33 & -2.32 & -28.99 & -28.99 & -17.77 & -20.57 & -28.57 & -20.11 \\
\cellcolor[HTML]{EFEFEF} & frozenui    & 0.01 & -0.16 & -9.01 & 0.67 & 0.35 & \cellcolor[HTML]{C0C0C0}0.87 & -7.04 & 0.71 \\
\cellcolor[HTML]{EFEFEF} & goby    & 0.43 & \cellcolor[HTML]{C0C0C0}0.98 & -1.05 & -0.61 & \cellcolor[HTML]{C0C0C0}0.99 & 0.75 & -0.74 & 0.57 \\
\cellcolor[HTML]{EFEFEF} & gosec    & -0.17 & -3.02 & \cellcolor[HTML]{C0C0C0}0.99 & 0.33 & 0.23 & \cellcolor[HTML]{C0C0C0}0.99 & 0.3 & 0.03 \\
\cellcolor[HTML]{EFEFEF} & gradle-play-publisher    & -1.13 & 0.43 & 0.81 & 0.13 & \cellcolor[HTML]{C0C0C0}0.96 & 0.89 & 0.58 & 0.77 \\
\cellcolor[HTML]{EFEFEF} & HackingWithSwift    & 0.48 & 0.39 & 0.57 & 0.52 & \cellcolor[HTML]{C0C0C0}0.67 & \cellcolor[HTML]{C0C0C0}0.67 & 0.53 & \cellcolor[HTML]{C0C0C0}0.64 \\
\cellcolor[HTML]{EFEFEF} & hadolint    & -0.17 & -2.39 & -7.51 & -2.51 & 0.14 & \cellcolor[HTML]{C0C0C0}0.88 & -6.77 & -0.39 \\
\cellcolor[HTML]{EFEFEF} & horizon    & -2.01 & -8.49 & \cellcolor[HTML]{C0C0C0}0.01 & -1.33 & -1.17 & -0.33 & -1.16 & -1.87 \\
\cellcolor[HTML]{EFEFEF} & ImageOptim-CLI    & 0.67 & 0.71 & \cellcolor[HTML]{C0C0C0}0.93 & 0.78 & 0.89 & \cellcolor[HTML]{C0C0C0}0.99 & \cellcolor[HTML]{C0C0C0}0.95 & \cellcolor[HTML]{C0C0C0}0.97 \\
\cellcolor[HTML]{EFEFEF} & javacpp    & -3.67 & -3.19 & -5.01 & -3.67 & \cellcolor[HTML]{C0C0C0}-1.56 & \cellcolor[HTML]{C0C0C0}-1.45 & -3.72 & -2.4 \\
\cellcolor[HTML]{EFEFEF} & KSCrash    & -1.33 & -1.18 & 0.01 & 0.01 & 0.86 & \cellcolor[HTML]{C0C0C0}0.98 & -0.03 & 0.36 \\
\cellcolor[HTML]{EFEFEF} & LayoutKit    & 0.33 & 0.74 & -3.67 & -2.78 & \cellcolor[HTML]{C0C0C0}0.96 & \cellcolor[HTML]{C0C0C0}0.96 & -1.61 & 0.23 \\
\cellcolor[HTML]{EFEFEF} & LeafPic    & -3.33 & -0.29 & \cellcolor[HTML]{C0C0C0}0.99 & 0.33 & 0.83 & \cellcolor[HTML]{C0C0C0}0.99 & 0.84 & 0.59 \\
\cellcolor[HTML]{EFEFEF} & lossless-cut    & 0.13 & 0.29 & 0.33 & 0.68 & \cellcolor[HTML]{C0C0C0}0.94 & \cellcolor[HTML]{C0C0C0}0.94 & 0.44 & 0.88 \\
\cellcolor[HTML]{EFEFEF} & material    & -21.22 & \cellcolor[HTML]{C0C0C0}-5.07 & -21.01 & -25.01 & -11.57 & -11.57 & -19.86 & -12.71 \\
\cellcolor[HTML]{EFEFEF} & mkdocs-material    & -5.67 & -21.56 & \cellcolor[HTML]{C0C0C0}0.99 & 0.33 & 0.67 & 0.67 & -1.71 & -3.45 \\
\cellcolor[HTML]{EFEFEF} & mobile-angular-ui    & -0.33 & 0.48 & \cellcolor[HTML]{C0C0C0}0.99 & -4.78 & \cellcolor[HTML]{C0C0C0}0.98 & \cellcolor[HTML]{C0C0C0}0.99 & 0.24 & -0.21 \\
\cellcolor[HTML]{EFEFEF} & moco    & 0.88 & 0.91 & 0.85 & 0.88 & \cellcolor[HTML]{C0C0C0}0.97 & \cellcolor[HTML]{C0C0C0}0.96 & \cellcolor[HTML]{C0C0C0}0.94 & \cellcolor[HTML]{C0C0C0}0.99 \\
\cellcolor[HTML]{EFEFEF} & neo    & 0.88 & 0.39 & 0.89 & 0.91 & 0.89 & \cellcolor[HTML]{C0C0C0}0.98 & 0.81 & 0.76 \\
\cellcolor[HTML]{EFEFEF} & open-location-code    & -3.33 & -2.65 & \cellcolor[HTML]{C0C0C0}0.01 & -1.67 & -1.23 & \cellcolor[HTML]{C0C0C0}0.01 & -0.52 & -0.92 \\
\cellcolor[HTML]{EFEFEF} & osv    & \cellcolor[HTML]{C0C0C0}-9.99 & -17.99 & -500.99 & -310.99 & \cellcolor[HTML]{C0C0C0}-10.32 & -56.32 & -455.06 & -88.99 \\
\cellcolor[HTML]{EFEFEF} & patroni    & 0.93 & 0.88 & 0.89 & 0.89 & 0.82 & \cellcolor[HTML]{C0C0C0}0.97 & 0.85 & 0.89 \\
\cellcolor[HTML]{EFEFEF} & pigeon-maps    & 0.67 & 0.62 & 0.01 & -0.99 & 0.83 & \cellcolor[HTML]{C0C0C0}0.87 & 0.02 & 0.52 \\
\cellcolor[HTML]{EFEFEF} & polaris-react    & \cellcolor[HTML]{C0C0C0}0.01 & -120.21 & -4.01 & -4.01 & -4.29 & -3.51 & -25.47 & -33.49 \\
\cellcolor[HTML]{EFEFEF} & polr    & -14.33 & -10.58 & -42.99 & -21.33 & \cellcolor[HTML]{C0C0C0}-3.54 & -16.49 & -44.59 & -18.75 \\
\cellcolor[HTML]{EFEFEF} & react-flip-move    & -0.33 & 0.11 & -5.01 & -1.01 & 0.65 & \cellcolor[HTML]{C0C0C0}0.71 & -3.16 & 0.25 \\
\cellcolor[HTML]{EFEFEF} & react-native-gesture-handler    & 0.87 & 0.28 & \cellcolor[HTML]{C0C0C0}0.99 & 0.01 & 0.67 & 0.62 & 0.87 & 0.78 \\
\cellcolor[HTML]{EFEFEF} & react-overdrive    & -2.01 & 0.09 & 0.51 & 0.51 & \cellcolor[HTML]{C0C0C0}0.76 & 0.71 & 0.3 & 0.55 \\
\cellcolor[HTML]{EFEFEF} & ring    & 0.67 & 0.36 & 0.19 & \cellcolor[HTML]{C0C0C0}0.99 & 0.39 & 0.78 & 0.37 & 0.75 \\
\cellcolor[HTML]{EFEFEF} & SCRecorder    & 0.17 & -1.23 & -1.01 & -0.51 & 0.71 & \cellcolor[HTML]{C0C0C0}0.92 & -0.65 & 0.16 \\
\cellcolor[HTML]{EFEFEF} & Solve-App-Store-Review-Problem    & -2.33 & 0.08 & -7.01 & -4.67 & 0.76 & \cellcolor[HTML]{C0C0C0}0.89 & -6.55 & -0.34 \\
\cellcolor[HTML]{EFEFEF} & sqldelight    & 0.72 & 0.55 & \cellcolor[HTML]{C0C0C0}0.97 & 0.79 & \cellcolor[HTML]{C0C0C0}0.95 & \cellcolor[HTML]{C0C0C0}0.98 & \cellcolor[HTML]{C0C0C0}0.93 & 0.9 \\
\cellcolor[HTML]{EFEFEF} & tabulator    & -1.44 & -1.99 & 0.22 & -1.15 & \cellcolor[HTML]{C0C0C0}0.99 & \cellcolor[HTML]{C0C0C0}0.99 & -0.1 & -0.07 \\
\cellcolor[HTML]{EFEFEF} & TheAmazingAudioEngine    & \cellcolor[HTML]{C0C0C0}0.98 & \cellcolor[HTML]{C0C0C0}0.98 & 0.83 & 0.01 & 0.81 & 0.93 & -1.26 & 0.6 \\
\cellcolor[HTML]{EFEFEF} & ts-jest    & 0.73 & 0.68 & 0.76 & 0.48 & 0.89 & \cellcolor[HTML]{C0C0C0}0.95 & 0.58 & 0.89 \\
\cellcolor[HTML]{EFEFEF} & vscode-cpptools    & \cellcolor[HTML]{C0C0C0}0.86 & 0.52 & 0.29 & 0.71 & 0.54 & 0.59 & 0.12 & 0.6 \\
\cellcolor[HTML]{EFEFEF} & vue-meta    & -0.17 & -3.64 & 0.01 & -4.17 & 0.39 & \cellcolor[HTML]{C0C0C0}0.51 & -1.43 & -1.28 \\
\cellcolor[HTML]{EFEFEF} & XcodeGen    & -2.14 & -6.04 & -1.86 & -1.19 & -1.33 & \cellcolor[HTML]{C0C0C0}0.33 & -1.36 & -1.48 \\
\multirow{-60}{*}{\cellcolor[HTML]{EFEFEF}\textbf{contemporary}} & XposedBridge    & \cellcolor[HTML]{C0C0C0}0.99 & 0.14 & 0.25 & 0.51 & 0.89 & \cellcolor[HTML]{C0C0C0}0.96 & 0.37 & 0.8 \\ \hline
\multicolumn{2}{|c|}{\multirow{-1}{*}{\cellcolor[HTML]{EFEFEF}\textbf{Win Time Count}}} & \multicolumn{1}{c|}{\textbf{9}} & \multicolumn{1}{c|}{\textbf{4}} & \multicolumn{1}{c|}{\textbf{13}} & \multicolumn{1}{c|}{\textbf{1}} & \multicolumn{1}{c|}{\textbf{22}} & \multicolumn{1}{c|}{\textbf{36}} & \multicolumn{1}{c|}{\textbf{3}} & \multicolumn{1}{c|}{\textbf{4}} \\ \hline
\end{tabular}

\end{adjustbox}
\end{table*}

\begin{table*}[!t]
\centering
\caption{SA (Standard Accuracy) of contemporary data sets (Part 2), \underline{higher} values are \underline{better}. Same format as Table~\ref{table:result_sa}; }
\label{table:result_sa_con_2}
\begin{adjustbox}{max width=0.9\textwidth}
\centering
\begin{tabular}{|c|l|cccccccc|}
\hline
\rowcolor[HTML]{EFEFEF} 
\multicolumn{2}{|c|}{\cellcolor[HTML]{EFEFEF}} & \multicolumn{4}{c|}{\cellcolor[HTML]{EFEFEF}\textbf{Scikit-Learn}} & \multicolumn{2}{c|}{\cellcolor[HTML]{EFEFEF}\textbf{Tuned}} & \multicolumn{2}{c|}{\cellcolor[HTML]{EFEFEF}\textbf{Other Methods}} \\ \cline{3-10} 
\multicolumn{2}{|c|}{\multirow{-2}{*}{\cellcolor[HTML]{EFEFEF}\textbf{Dataset}}} & \multicolumn{1}{c|}{\textbf{KNN}} & \multicolumn{1}{c|}{\textbf{SVR}} & \multicolumn{1}{c|}{\textbf{CART}} & \multicolumn{1}{c|}{\textbf{RF}} & \multicolumn{1}{c|}{\textbf{CART\_DE}} & \multicolumn{1}{c|}{\textbf{ROME}} & \multicolumn{1}{c|}{\textbf{ATLM}} & \multicolumn{1}{c|}{\textbf{LP4EE}} \\ \hline
\cellcolor[HTML]{EFEFEF} & aeron    & -24.17 & \cellcolor[HTML]{C0C0C0}-15.95 & -29.51 & -20.92 & -17.51 & \cellcolor[HTML]{C0C0C0}-13.07 & -28.28 & \cellcolor[HTML]{C0C0C0}-15.88 \\
\cellcolor[HTML]{EFEFEF} & alasql    & -0.22 & 0.77 & 0.33 & -3.04 & 0.87 & \cellcolor[HTML]{C0C0C0}0.99 & 0.14 & 0.18 \\
\cellcolor[HTML]{EFEFEF} & android-advancedrecyclerview    & 0.26 & 0.87 & \cellcolor[HTML]{C0C0C0}0.99 & 0.78 & 0.71 & 0.88 & 0.91 & 0.93 \\
\cellcolor[HTML]{EFEFEF} & Android-Image-Cropper    & -7.22 & \cellcolor[HTML]{C0C0C0}-0.08 & -14.67 & -4.78 & -0.85 & \cellcolor[HTML]{C0C0C0}-0.11 & -8.97 & -0.77 \\
\cellcolor[HTML]{EFEFEF} & AndroidPicker    & -1.33 & \cellcolor[HTML]{C0C0C0}0.02 & -2.01 & -1.17 & -1.15 & -1.15 & -1.73 & -0.96 \\
\cellcolor[HTML]{EFEFEF} & angular-seed    & -1.67 & -6.69 & 0.67 & 0.11 & -1.77 & \cellcolor[HTML]{C0C0C0}0.58 & -0.71 & -0.94 \\
\cellcolor[HTML]{EFEFEF} & BackstopJS    & \cellcolor[HTML]{C0C0C0}0.33 & \cellcolor[HTML]{C0C0C0}0.31 & -0.33 & -0.37 & 0.06 & 0.16 & -0.85 & 0.1 \\
\cellcolor[HTML]{EFEFEF} & brave-browser    & 0.58 & -0.44 & -0.45 & 0.85 & \cellcolor[HTML]{C0C0C0}0.92 & \cellcolor[HTML]{C0C0C0}0.97 & -1.29 & 0.69 \\
\cellcolor[HTML]{EFEFEF} & cachecloud    & -0.78 & \cellcolor[HTML]{C0C0C0}0.99 & -8.67 & -4.56 & -1.36 & -0.27 & -7.15 & -0.81 \\
\cellcolor[HTML]{EFEFEF} & caprine    & 0.67 & 0.75 & 0.63 & 0.51 & \cellcolor[HTML]{C0C0C0}0.99 & \cellcolor[HTML]{C0C0C0}0.98 & 0.75 & \cellcolor[HTML]{C0C0C0}0.93 \\
\cellcolor[HTML]{EFEFEF} & cfssl    & 0.67 & 0.27 & 0.51 & 0.83 & \cellcolor[HTML]{C0C0C0}0.88 & \cellcolor[HTML]{C0C0C0}0.92 & 0.64 & \cellcolor[HTML]{C0C0C0}0.9 \\
\cellcolor[HTML]{EFEFEF} & clappr    & -0.51 & -8.31 & 0.75 & -8.01 & -2.63 & \cellcolor[HTML]{C0C0C0}0.94 & -2.18 & -3.07 \\
\cellcolor[HTML]{EFEFEF} & community-edition    & 0.78 & -2.35 & 0.67 & 0.56 & \cellcolor[HTML]{C0C0C0}0.99 & \cellcolor[HTML]{C0C0C0}0.99 & 0.3 & 0.42 \\
\cellcolor[HTML]{EFEFEF} & core    & 0.24 & -0.85 & -2.82 & 0.31 & -0.54 & \cellcolor[HTML]{C0C0C0}0.45 & -0.89 & 0.25 \\
\cellcolor[HTML]{EFEFEF} & cpprestsdk    & -3.08 & -0.62 & 0.25 & 0.51 & 0.63 & \cellcolor[HTML]{C0C0C0}0.97 & 0.27 & 0.64 \\
\cellcolor[HTML]{EFEFEF} & Dexie.js    & -7.01 & -6.15 & -3.51 & -4.33 & \cellcolor[HTML]{C0C0C0}-2.79 & \cellcolor[HTML]{C0C0C0}-2.79 & -5.25 & -3.55 \\
\cellcolor[HTML]{EFEFEF} & diesel    & 0.67 & 0.75 & -2.04 & 0.13 & \cellcolor[HTML]{C0C0C0}0.98 & \cellcolor[HTML]{C0C0C0}0.98 & -1.55 & 0.86 \\
\cellcolor[HTML]{EFEFEF} & discord.js    & 0.65 & -0.44 & 0.48 & 0.57 & \cellcolor[HTML]{C0C0C0}0.98 & \cellcolor[HTML]{C0C0C0}0.98 & 0.6 & 0.58 \\
\cellcolor[HTML]{EFEFEF} & documentation    & 0.75 & -0.48 & 0.75 & 0.51 & \cellcolor[HTML]{C0C0C0}0.91 & \cellcolor[HTML]{C0C0C0}0.91 & 0.49 & 0.54 \\
\cellcolor[HTML]{EFEFEF} & EarlGrey    & 0.01 & 0.32 & 0.01 & 0.33 & \cellcolor[HTML]{C0C0C0}0.57 & \cellcolor[HTML]{C0C0C0}0.58 & 0.05 & 0.49 \\
\cellcolor[HTML]{EFEFEF} & EIPs    & \cellcolor[HTML]{C0C0C0}0.92 & 0.79 & 0.56 & 0.55 & 0.74 & 0.74 & 0.63 & 0.77 \\
\cellcolor[HTML]{EFEFEF} & error-prone    & -14.33 & -25.22 & 0.01 & 0.01 & -1.79 & \cellcolor[HTML]{C0C0C0}0.51 & -4.59 & -6.4 \\
\cellcolor[HTML]{EFEFEF} & evil-icons    & 0.53 & 0.63 & 0.81 & 0.73 & \cellcolor[HTML]{C0C0C0}0.88 & \cellcolor[HTML]{C0C0C0}0.91 & 0.73 & 0.76 \\
\cellcolor[HTML]{EFEFEF} & exceljs    & -0.75 & -0.41 & 0.75 & 0.83 & \cellcolor[HTML]{C0C0C0}0.95 & \cellcolor[HTML]{C0C0C0}0.99 & 0.6 & 0.7 \\
\cellcolor[HTML]{EFEFEF} & fission    & 0.48 & 0.88 & 0.84 & 0.79 & \cellcolor[HTML]{C0C0C0}0.99 & \cellcolor[HTML]{C0C0C0}0.99 & 0.82 & \cellcolor[HTML]{C0C0C0}0.95 \\
\cellcolor[HTML]{EFEFEF} & flannel    & 0.88 & \cellcolor[HTML]{C0C0C0}0.96 & 0.79 & 0.76 & \cellcolor[HTML]{C0C0C0}0.98 & \cellcolor[HTML]{C0C0C0}0.99 & 0.84 & 0.89 \\
\cellcolor[HTML]{EFEFEF} & go-git    & -2.56 & -5.93 & \cellcolor[HTML]{C0C0C0}0.33 & -0.78 & 0.15 & 0.15 & -1.04 & -1.48 \\
\cellcolor[HTML]{EFEFEF} & goreleaser    & -0.26 & -1.65 & \cellcolor[HTML]{C0C0C0}0.69 & -0.53 & \cellcolor[HTML]{C0C0C0}0.69 & \cellcolor[HTML]{C0C0C0}0.71 & 0.09 & 0.06 \\
\cellcolor[HTML]{EFEFEF} & haven    & -0.18 & 0.77 & 0.91 & 0.67 & 0.93 & \cellcolor[HTML]{C0C0C0}0.99 & 0.77 & 0.78 \\
\cellcolor[HTML]{EFEFEF} & hospitalrun-frontend    & -1.08 & -1.98 & -0.15 & -1.92 & -0.24 & \cellcolor[HTML]{C0C0C0}0.13 & -0.58 & -0.56 \\
\cellcolor[HTML]{EFEFEF} & HTextView    & -0.33 & \cellcolor[HTML]{C0C0C0}0.93 & -0.24 & -1.01 & \cellcolor[HTML]{C0C0C0}0.92 & \cellcolor[HTML]{C0C0C0}0.92 & 0.18 & 0.52 \\
\cellcolor[HTML]{EFEFEF} & humhub    & \cellcolor[HTML]{C0C0C0}-19.67 & -59.01 & -41.01 & -32.01 & -29.88 & -29.88 & -48.78 & -32.92 \\
\cellcolor[HTML]{EFEFEF} & huxpro.github.io    & \cellcolor[HTML]{C0C0C0}0.99 & -3.41 & -2.01 & -5.67 & -1.93 & -7.77 & -5.9 & -5.77 \\
\cellcolor[HTML]{EFEFEF} & incubator-tvm    & -1.09 & -8.83 & \cellcolor[HTML]{C0C0C0}-0.43 & -1.62 & -1.36 & -1.36 & -2.25 & -2.98 \\
\cellcolor[HTML]{EFEFEF} & is-thirteen    & 0.51 & 0.45 & 0.41 & \cellcolor[HTML]{C0C0C0}0.57 & \cellcolor[HTML]{C0C0C0}0.56 & \cellcolor[HTML]{C0C0C0}0.56 & 0.32 & \cellcolor[HTML]{C0C0C0}0.62 \\
\cellcolor[HTML]{EFEFEF} & jellyfin    & \cellcolor[HTML]{C0C0C0}-12.46 & -15.59 & -16.97 & -17.25 & -16.66 & -16.61 & -21.98 & -19.03 \\
\cellcolor[HTML]{EFEFEF} & kanboard    & 0.44 & -3.87 & \cellcolor[HTML]{C0C0C0}0.83 & 0.22 & 0.67 & 0.23 & -0.05 & -0.61 \\
\cellcolor[HTML]{EFEFEF} & kingshard    & 0.61 & \cellcolor[HTML]{C0C0C0}0.89 & -0.33 & -0.44 & \cellcolor[HTML]{C0C0C0}0.94 & \cellcolor[HTML]{C0C0C0}0.94 & -3.21 & 0.71 \\
\cellcolor[HTML]{EFEFEF} & lucida    & 0.67 & -1.55 & \cellcolor[HTML]{C0C0C0}0.99 & 0.83 & 0.61 & 0.75 & 0.46 & 0.28 \\
\cellcolor[HTML]{EFEFEF} & moon    & 0.78 & 0.14 & 0.33 & 0.67 & 0.53 & \cellcolor[HTML]{C0C0C0}0.98 & 0.57 & 0.8 \\
\cellcolor[HTML]{EFEFEF} & mybatis-generator-gui    & -1.33 & -2.62 & -0.41 & 0.53 & \cellcolor[HTML]{C0C0C0}0.67 & 0.43 & -0.37 & -0.33 \\
\cellcolor[HTML]{EFEFEF} & ngx-bootstrap    & -2.22 & -8.32 & 0.33 & 0.33 & 0.33 & \cellcolor[HTML]{C0C0C0}0.94 & -0.45 & -0.84 \\
\cellcolor[HTML]{EFEFEF} & or-tools    & \cellcolor[HTML]{C0C0C0}-10.17 & -19.14 & \cellcolor[HTML]{C0C0C0}-9.99 & -15.01 & -20.92 & \cellcolor[HTML]{C0C0C0}-9.81 & -12.58 & -13.23 \\
\cellcolor[HTML]{EFEFEF} & oss-fuzz    & \cellcolor[HTML]{C0C0C0}-7.73 & -10.91 & -9.21 & -9.21 & -9.51 & \cellcolor[HTML]{C0C0C0}-7.45 & -10.71 & \cellcolor[HTML]{C0C0C0}-7.94 \\
\cellcolor[HTML]{EFEFEF} & places    & 0.85 & 0.85 & 0.89 & \cellcolor[HTML]{C0C0C0}0.96 & \cellcolor[HTML]{C0C0C0}0.95 & \cellcolor[HTML]{C0C0C0}0.98 & \cellcolor[HTML]{C0C0C0}0.92 & \cellcolor[HTML]{C0C0C0}0.94 \\
\cellcolor[HTML]{EFEFEF} & pulsar    & -1.71 & -8.56 & \cellcolor[HTML]{C0C0C0}0.01 & -1.44 & -1.21 & -0.44 & -2.33 & -2.6 \\
\cellcolor[HTML]{EFEFEF} & react-autosuggest    & 0.01 & -4.02 & \cellcolor[HTML]{C0C0C0}0.99 & 0.01 & 0.82 & 0.86 & -0.06 & -0.45 \\
\cellcolor[HTML]{EFEFEF} & react-map-gl    & 0.14 & 0.54 & 0.57 & 0.05 & \cellcolor[HTML]{C0C0C0}0.88 & \cellcolor[HTML]{C0C0C0}0.88 & 0.41 & 0.65 \\
\cellcolor[HTML]{EFEFEF} & react-native-paper    & 0.77 & \cellcolor[HTML]{C0C0C0}0.91 & -0.09 & 0.69 & 0.62 & 0.69 & 0.27 & 0.76 \\
\cellcolor[HTML]{EFEFEF} & redex    & -15.01 & -18.89 & -11.01 & -16.67 & -10.92 & \cellcolor[HTML]{C0C0C0}-8.83 & -15.07 & -12.79 \\
\cellcolor[HTML]{EFEFEF} & resilience4j    & -3.56 & 0.48 & 0.33 & 0.67 & 0.64 & \cellcolor[HTML]{C0C0C0}0.79 & 0.21 & 0.58 \\
\cellcolor[HTML]{EFEFEF} & SCLAlertView-Swift    & 0.67 & -0.95 & 0.01 & 0.67 & 0.59 & \cellcolor[HTML]{C0C0C0}0.89 & -0.02 & 0.53 \\
\cellcolor[HTML]{EFEFEF} & sdk    & -12.83 & -14.29 & -12.26 & -13.65 & -12.26 & \cellcolor[HTML]{C0C0C0}-10.26 & -12.6 & -11.3 \\
\cellcolor[HTML]{EFEFEF} & single-spa    & 0.11 & -0.39 & 0.67 & 0.67 & 0.54 & \cellcolor[HTML]{C0C0C0}0.99 & 0.55 & 0.64 \\
\cellcolor[HTML]{EFEFEF} & Squirrel.Windows    & 0.67 & 0.77 & 0.33 & 0.78 & \cellcolor[HTML]{C0C0C0}0.91 & 0.82 & 0.48 & 0.76 \\
\cellcolor[HTML]{EFEFEF} & thingsboard    & 0.57 & 0.78 & 0.44 & 0.67 & \cellcolor[HTML]{C0C0C0}0.91 & 0.87 & -0.42 & 0.84 \\
\cellcolor[HTML]{EFEFEF} & TranslationPlugin    & -0.22 & 0.77 & 0.24 & -1.05 & \cellcolor[HTML]{C0C0C0}0.84 & -0.43 & -0.05 & -0.29 \\
\cellcolor[HTML]{EFEFEF} & translations    & \cellcolor[HTML]{C0C0C0}0.44 & 0.31 & -9.01 & -3.67 & \cellcolor[HTML]{C0C0C0}0.45 & 0.06 & -5.97 & -0.45 \\
\cellcolor[HTML]{EFEFEF} & vue-multiselect    & -7.33 & -7.45 & -6.01 & -12.01 & \cellcolor[HTML]{C0C0C0}-5.46 & \cellcolor[HTML]{C0C0C0}-5.46 & -11.3 & -8.07 \\
\multirow{-60}{*}{\cellcolor[HTML]{EFEFEF}\textbf{contemporary}} & z3    & -10.73 & -21.25 & -5.51 & -6.36 & -8.67 & \cellcolor[HTML]{C0C0C0}-4.92 & -11.75 & -8.33 \\ \hline
\multicolumn{2}{|c|}{\multirow{-1}{*}{\cellcolor[HTML]{EFEFEF}\textbf{Win Time Count}}} & \multicolumn{1}{c|}{\textbf{8}} & \multicolumn{1}{c|}{\textbf{9}} & \multicolumn{1}{c|}{\textbf{9}} & \multicolumn{1}{c|}{\textbf{2}} & \multicolumn{1}{c|}{\textbf{25}} & \multicolumn{1}{c|}{\textbf{40}} & \multicolumn{1}{c|}{\textbf{1}} & \multicolumn{1}{c|}{\textbf{7}} \\ \hline
\end{tabular}

\end{adjustbox}
\end{table*}

\section{Results}

In this section, we present the experimental results. To answer the questions raised in Section~\ref{sect:intro}, we conduct the following experiments:
\bi
\item Compare performance of ROME with other methods on COCOMO-style data, classic effort data and contemporary data sets collected from Github.
\item Look into the internal structure of ROME and count the feature node in the tree it built.
\ei

{\bf RQ1: Is effort estimation effective for classic waterfall and contemporary projects?}

To find if   effort estimation method is effective, we ran ROME on both classic waterfall data sets and contemporary data sets. The performance value of classic waterfall data sets, in terms of MRE, is shown in Table~\ref{table:result_mre}. Recall that
Sarro et al. argued that effective software projects have predictions of effort lie   0.3 and 0.4 of the actual value~\cite{sarro2016multi}.
As can be observed, ROME obtained the lowest MRE value less than 0.40 (in 8 out of all 12 cases).
Also, in terms of applicability to contemporary methods (shown in Table~\ref{table:result_mre_con_1} and Table~\ref{table:result_mre_con_2}), 
it is significant to note that many of the MREs seen in the
  contemporary projects are under 0.30.  That is, with these   results, we can recommend ROME to the current practice, especially for the current contemporary projects. Overall:

\begin{result}{1}
Effort estimation is effective on both classic waterfall projects and contemporary projects.
 \end{result}





\begin{table}
\small
 \caption{The frequency of each treatment seen to be best in classic data sets. }\label{tbl:methodsrk}
\centering \begin{tabular}{clc} \\
  {\textbf{Rank}}& \textbf{Method} & \textbf{Win Times (percentage)}\\
  \hline
    1 &      ROME           &    21/24 (87.5\%)\\
    2 &      CART\_DE       &    15/24 (62.5\%)\\
    3 &      LP4EE          &    9/24 (37.5\%)\\
    4 &      KNN            &    2/24 (8.3\%)\\
    4 &      RF             &    2/24 (8.3\%)\\
    5 &      CART           &    1/24 (4.2\%)\\
    6 &      SVR            &    0/24 (0\%)\\
    6 &      ATLM           &    0/24 (0\%)\\
    \hline   
\end{tabular}
\end{table}

\begin{table}
\small
 \caption{The frequency of each treatment seen to be best in contemporary data sets. }\label{tbl:methodsrk_con}
\centering \begin{tabular}{clc} \\
  {\textbf{Rank}}& \textbf{Method} & \textbf{Win Times (percentage)}\\
  \hline
    1 &      ROME           &    148/240 (61.7\%)\\
    2 &      CART\_DE       &    116/240 (48.3\%)\\
    3 &      CART           &    44/240 (18.3\%)\\
    4 &      KNN            &    32/240 (13.3\%)\\
    5 &      SVR            &    24/240 (10\%)\\
    6 &      LP4EE          &    21/240 (8.8\%)\\
    7 &      RF             &    8/240 (3.3\%)\\
    8 &      ATLM           &    5/240 (2.1\%)\\
    \hline   
\end{tabular}
\end{table}



In terms of the practicality of effort estimation research, this is a landmark result since it means that decades of research into effort estimation of classic waterfall projects can now be applied to contemporary software systems.

{\bf RQ2: Does ROME have better performance than existing estimation methods?}

To answer this question, we ran ROME  and the other baseline methods {\it LP4EE}, {\it ATLM}, {\it KNN}, {\it SVR}, {\it CART}, {\it RF}, on classic waterfall data sets and contemporary data sets. MRE scores for all our methods are shown in Table~\ref{table:result_mre}, Table~\ref{table:result_mre_con_1} and Table~\ref{table:result_mre_con_2}. 
While SA scores are shown in Table~\ref{table:result_sa}, Table~\ref{table:result_sa_con_1} and Table~\ref{table:result_sa_con_2}. Note the COCOMO-II is only applied to the COCOMO data sets (since the other data sets do not have the features needed
by COCOMO).

In those tables,  each row shows results from a different data set.
 For each row, the gray cells show the results that are statistically significantly  better than anything else on that row (as judged by a Scott-Knot bootstrap test plus an  A12 effect size test).
If multiple treatments tied for ``best'', then there will be multiple gray cells in a row, better methods have more gray cells. 
 Table~\ref{tbl:methodsrk} and Table~\ref{tbl:methodsrk_con} tallies the gray cells counts for all methods.

 From the tallies of Table~\ref{tbl:methodsrk} and Table~\ref{tbl:methodsrk_con}, we conclude that RF and ATLM most often perform worse than anything else. While KNN, SVR, CART (untuned) and LP4EE does better than the previous two in contemporary data sets, they are not competitive against the tuned methods (CART, tuned by DE or FLASH). In classic data set, LP4EE does better than other untuned methods but still not competitive against the tuned ones.
 As to DE tuning CART, it performs better than all untuned methods, but not as good as the method with powered by sequential model-based optimization, ROME, in both classic and contemporary data sets. In summary:
 \begin{result}{2}
ROME generate better estimates than other methods in most cases.
 \end{result}





\noindent{\bf RQ3: When we have new effort data sets, what configurations  to use for effort estimation tasks?}

\begin{figure*}[!t]
\begin{center}
\renewcommand{\baselinestretch}{0.75}


\begin{tabular}{ r|c|c|c|c|c|c}

~ & \%max\_features & max\_depth & min\_sample\_split & min\_samples\_leaf  \\  
~ & (selected at random; & (of trees) & (continuation & (termination \\ 
~ & 100\% means ``use all'') &   & criteria) & criteria)   \\\cline{2-7} 
~ &
\makecell[l]{
\ 25\%\ 50\%\ 75\%\ 100\%} &
\makecell[l]{
\ $\leq$03 \ $\leq$06 \ $\leq$09 \ $\leq$12} &
\makecell[l]{
\ $\leq$5 \ $\leq$10 \ $\leq$15 \ $\leq$20} &
\makecell[l]{
\ $\leq$03 \ $\leq$06 \ $\leq$09 \ $\leq$12} 
 \\
\hline

cocomo10
&\dbox{23}\dbox{38}\dbox{18}\dbox{21}
&\dbox{42}\dbox{45}\dbox{11}\dbox{02}
&\wbox{85}\dbox{11}\dbox{04}\dbox{00}
&\wbox{79}\dbox{12}\dbox{06}\dbox{03}
\\
cocomo81
&\dbox{26}\dbox{33}\dbox{18}\dbox{23}
&\wbox{52}\dbox{22}\dbox{18}\dbox{08}
&\wbox{73}\dbox{25}\dbox{02}\dbox{00}
&\wbox{78}\dbox{17}\dbox{04}\dbox{01}
\\
nasa93
&\dbox{31}\dbox{27}\dbox{28}\dbox{24}
&\dbox{47}\dbox{29}\dbox{18}\dbox{06}
&\wbox{55}\dbox{21}\dbox{11}\dbox{13}
&\wbox{53}\dbox{27}\dbox{14}\dbox{06}
\\
contemporary
&\dbox{34}\dbox{41}\dbox{12}\dbox{13}
&\dbox{30}\wbox{52}\dbox{15}\dbox{03}
&\wbox{74}\dbox{13}\dbox{09}\dbox{04}
&\wbox{68}\dbox{25}\dbox{05}\dbox{02}
\\\end{tabular}

\mbox{KEY: \colorbox{black!10}{\bf 10}\colorbox{black!20}{\bf 20}\colorbox{black!30}{\bf 30}\colorbox{black!40}{\bf 40}\colorbox{black!50}{\bf \textcolor{white}{50}}\colorbox{black!60}{\bf \textcolor{white}{60}}\colorbox{black!70}{\bf \textcolor{white}{70}}\colorbox{black!80}{\bf \textcolor{white}{80}}\colorbox{black!90}{\bf \textcolor{white}{90}}\colorbox{black}{\bf \textcolor{white}{100}}}\%
\end{center}
\caption{Tunings discovered by hyperparameter selections
(CART+FLASH, MRE results). 
Cells in this table show the percent of times a particular choice was made. White text on black denotes choices made in more than 50\% of tunings.
}\label{fig:para_dist}
\end{figure*}

 When we discuss this work with our industrial colleagues, they want to know ``the bottom line''; i.e. what they should use or, at the very least, what they should not use.  
If the hyperparameter tunings for effort estimators found by this paper were nearly always the same, then this study
could conclude by recommending better values for default settings. This would
be a most promising result since, in future when new data arrives, the complexities of tuning in ROME framework would not be needed.

Unfortunately, this turns out not to be the case.
Figure~\ref{fig:para_dist} shows the percent frequencies with which
some tuning decision appears in our experiments
(this table uses results from FLASH tuning CART since, as shown below,
this usually leads to best MRE results).
Note that in those results it is not true that across most data sets there is a setting that is usually selected
(though min\_samples\_leaf less than 3 is often a  popular setting).
Accordingly, from Figure~\ref{fig:para_dist}, we concludes that there is much variations of the best tunings. 

This finding is quite aligned with Fu et al.~\cite{Fu2016TuningFS}, where for software defect predictors, no best tunings for all tasks. Therefore, we always prefer to have a fast hyperparameter tuning technique to quickly find the best tuning for the current tasks. Our ROME framework is such of tool to use.

Since there are no ``best'' default settings for all, based on the results of  Table~\ref{tbl:methodsrk} and Table~\ref{tbl:methodsrk_con}, 
for similar effort estimation tasks, we say:

\begin{result}{3}
There is no clear pattern in what configurations  are needed. Hence,  model optimization needs to be repeated for each new data set.
 \end{result}

{\bf RQ4: When we apply ROME on effort data sets, can it help us to find the most important features of the data?}

When CART's tuning parameters were described in \S\ref{sec:algo}, it was observed  that when CART  is run multiple times (with different hyperparameters)
then it can be used to gauge the value of using a particular feature.

\begin{figure}[!t]
\centerline{\includegraphics[width=0.5\textwidth]{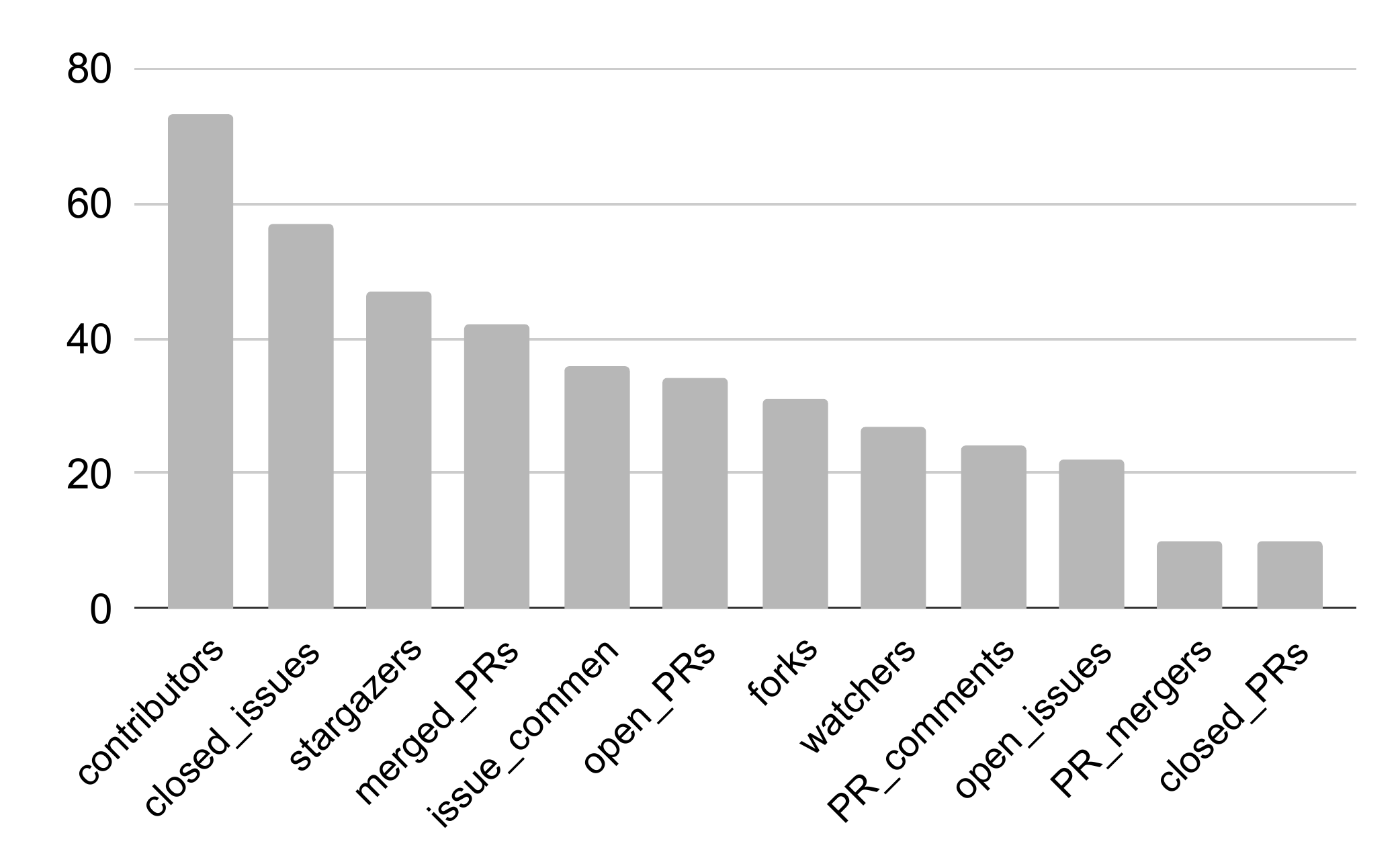}}
\caption{Selected feature count on contemporary data sets (for SA)}    
\label{fig:github_feature_sa}
\vspace{-4mm}
\end{figure}

\begin{figure}[!t]
\centerline{\includegraphics[width=0.5\textwidth]{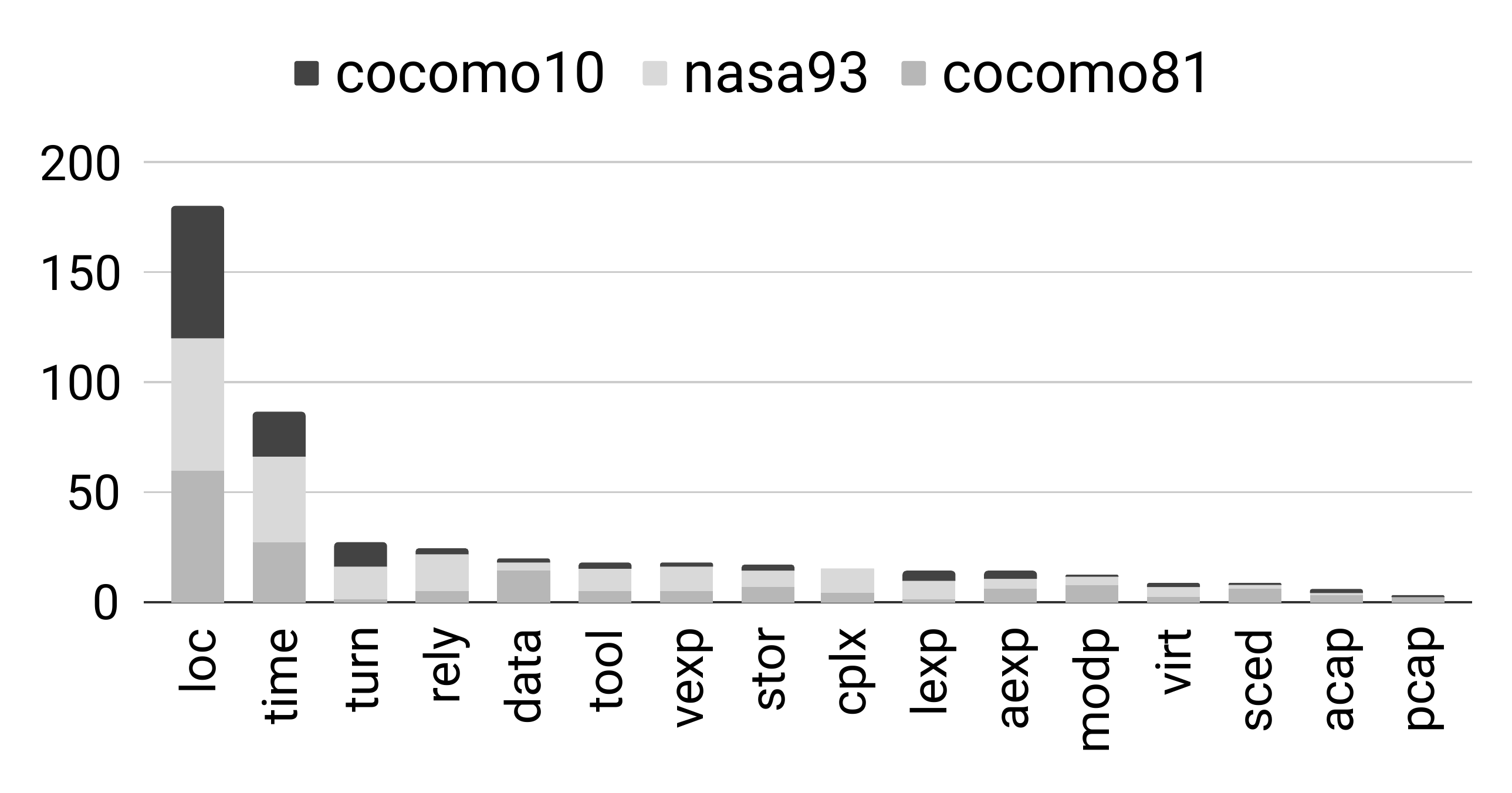}}
\caption{Selected feature count on COCOMO data sets (for SA)}    
\label{fig:cocomo_feature_sa}
\vspace{-4mm}
\end{figure}

Figure~\ref{fig:github_feature_sa} and Figure~\ref{fig:cocomo_feature_sa} show counts of how often a feature appeared in the   trees found by ROME  from the above experiments. Here, we  only show data from the classic COCOMO and contemporary Github projects since the classic non-COCOMO data sets
all use different features.

In Figure~\ref{fig:github_feature_sa}, the maximum number of times a feature can appear is 120 times (from 120 Github repositories).
One attribute ``contributors'' appears very frequently but it is not often  picked just by itself (we know this from the 
max\_depth results of Figure~\ref{fig:para_dist} where more often than not, CART used trees that held more than three features).
But as to what other features were combined with contributors, that is clear. Looking at the closed\_issues, stargazers, merged\_PRs, etc. results of Figure~\ref{fig:github_feature_sa}, we see that every other feature got used, sometimes.

A similar pattern appears in Figure~\ref{fig:cocomo_feature_sa}. In this figure, the maximum number of times a feature can appear is 180 (3 data sets, 3 way cross-validation, 20 repeats).
Once again, a size attribute (LOC) appears very frequently. But just as before, we see that   every other feature got used, sometimes.
Hence we say: 

\begin{result}{4}
There are no ``best'' set of effort estimation features since each project uses these features in a different way
\end{result}
As mentioned in the introduction,
the results from {\bf RQ3} and {\bf RQ4}  clearly deprecate the use of off-the-shelf estimation tools.   Practitioners
should use tools like ROME to find the features/modeling options that work best for their local data.

\section{Threats to Validity}\label{sect:threats}

The design of this study may have several validity threats~\cite{feldt2010validity}. The following issues should be considered to avoid jeopardizing conclusions made from this work:

\textbf{Internal Bias:} Many of  our methods contain stochastic random operators. To reduce the bias from random operators, we 
repeated our experiment in 20 times and applied statistical tests to remove spurious distinctions.

\textbf{Parameter Bias:} For other studies, this is a significant question
since (as shown above) the settings to the control parameters of the learners
can have a positive effect on the efficacy of the estimation.
That said, recall that much of the technology of this paper concerned methods to explore the space of possible parameters. Hence we assert that this study suffers much less parameter bias than other studies.
 

\textbf{Sampling Bias:} While we tested ROME on both old COCOMO-Style data sets, classic effort data sets and newly collected open source data sets, it would be inappropriate to conclude that ROME tuning  always perform better than others methods for other data sets.
As researchers, what we can do to mitigate this problem is to carefully document our methods, publish our tools as open source software packages,
and support the research community as they try to repeat/improve/refute our results on a broader set of data.


Another sampling bias comes from our choice of effort estimation technologies. Here, we compared ROME against technologies that are often seen  in the effort estimation literature.  We also took care to include in our comparisons two new and prominent methods recently published in TOSEM. But even with all that, this study has not explored all the effort estimation methods seen in the recent literature.
To some extent, that was because no single paper can explore
all algorithms. But also, sometimes we choose not to explore certain
algorithms since they are out-of-scope for this study.
  For example, apart from LP4EE, Sarro et al. also offer another estimation method based on
genetic algorithms called CoGEE~\cite{sarro2016multi}. That
tool optimizes for multiple goals so it would not be a fair
comparison to the tools used here (in defense of that decision,
we note that the authors do not compare LP4EE to CoGEE in their
TOSEM'18 paper).

\section{Conclusions and Future Work} \label{sect:conclusion}

Classic waterfall projects and contemporary projects have difference in their developing process. Effort estimation methods need to support both of these projects.
For something as complex as the effort estimation of modern software projects, no single method works best. Instead, best results come from trying out a large number of candidate methods. 

Sequential model-based optimization (SMO) is an effective way to explore a range of configuration options for effort estimation. 
Our sequential optimizer came from research into software configuration. One of the new insights that leads to this paper was that ``configuration'' is a synonym for ``hyperparameter optimziation''. Hence, hyperparameter-optimization-via-configuration tools
has not previously been explored in the literature.  
Also, prior to this paper, such optimizers have not been used
for   effort estimation.

When this optimizer was applied to  1161 classic waterfall projects and 120 contemporary projects we found that:
\bi
\item {\bf RQ1:} we could successfully apply the same optimization method to classic and contemporary projects. This is a significant  result since it means that decades of effort estimation research  can now be applied to contemporary systems.
\item {\bf RQ2:} those optimizations yield better estimates than other methods studied here.
\item {\bf RQ3, RQ4:} different data sets need different hyperparameter optimizations and use different features. 
This   means that we should  deprecate the use of off-the-shelf estimation tools.  Practitioners should use tools like ROME to find the features/modeling options that work best for their local data.
\ei
To the best of our knowledge, this is the largest effort estimation experiment yet reported.

As to future work, there is much   to do.
Clearly, we need to try other learners
(e.g. neural nets, Bayesian learners or gradient boosting tree)
and other optimizers (e.g. SMAC~\cite{DBLP:journals/corr/abs-1709-04636} or
vZ~\cite{bjorner2015nuz}).

Also, now that we can use Github data for effort estimation, it is time to scale this analysis to the large number of projects available at that source.  In the study of this paper, our 
{\bf RQ3, RQ4} results found no stability in the features used or hyperparameter options selected. We conjecture that such stable conclusions may exist-- if we look at much more project data.

More generally, in the study of effort estimation, most prior work only focus on comparisons of new estimation methods, but very less studies comparing latest technique with old classic models (e.g. COCOMO). Given the results of this paper, it is now important to validate newly proposed methods against  different type of effort project data sets (e.g. Waterfall and Contemporary). Further, if we are mining current Github projects, we might be able to use the methods of this paper to go beyond mere effort estimation to look better predict for  other measures of project health (e.g. number of new contributors  each month).



\bibliographystyle{IEEEtran}
\bibliography{reference}

\end{document}